\documentclass[apj]{emulateapj}
\usepackage{mathptmx}

\newcommand{\cf}   {{\ifmmode{C_f}                       \else{$C_{f}$}\fi}}
\newcommand{\zem}  {{\ifmmode{z_{em}}                    \else{$z_{em}$}\fi}}
\newcommand{\zabs} {{\ifmmode{z_{abs}}                   \else{$z_{abs}$}\fi}}
\newcommand{\kms}  {{\ifmmode{{\rm km~s}^{-1}}           \else{km~s$^{-1}$}\fi}}
\newcommand{\delv} {{\ifmmode{\Delta v}                  \else{$\Delta v$}\fi}}
\newcommand{\vmax} {{\ifmmode{v_{\rm max}}               \else{$v_{\rm max}$}\fi}}
\newcommand{\voff} {{\ifmmode{\upsilon_{\rm shift}}      \else{$\upsilon_{\rm shift}$}\fi}}
\newcommand{\wrest}{{\ifmmode{W_{\rm rest}}              \else{$W_{\rm rest}$}\fi}}
\newcommand{\cmm}  {{\ifmmode{{\rm cm}^{-2}}             \else{cm$^{-2}$}\fi}}
\newcommand{\cmmm} {{\ifmmode{{\rm cm}^{-3}}             \else{cm$^{-3}$}\fi}}
\newcommand{\nh}   {{\ifmmode{N_{\rm H}}                 \else{$N_{\rm H}$}\fi}}
\newcommand{\nhgal}{{\ifmmode{N_{\rm H}^{\rm Gal}}       \else{$N_{\rm H}^{\rm Gal}$}\fi}}
\newcommand{\aox}  {{\ifmmode{\alpha_{\rm ox}}           \else{$\alpha_{\rm ox}$}\fi}}
\newcommand{\aoxpr}{{\ifmmode{\alpha_{\rm ox}^{\prime}}  \else{$\alpha_{\rm ox}^{\prime}$}\fi}}
\newcommand{\daox} {{\ifmmode{\Delta\alpha_{\rm ox}}     \else{$\Delta\alpha_{\rm ox}$}\fi}}

\newcommand{\sang}  {\buildrel {}_{\circ} \over {\rm A}}
\newcommand{\sAuv}  {A\raise 0.3 em \hbox{$_{\scriptscriptstyle 2500\;\sang}$}}
\newcommand{\Auv}   {\ifmmode{A\raise 0.3 em \hbox{$_{\scriptstyle 2500\;\sang}$}}
                     \else{$A\raise 0.3 em \hbox{$_{\scriptstyle 2500\;\sang}$}$} \fi}
\newcommand{\luv}   {\ifmmode{\ell\raise 0.3 em \hbox{$_{\scriptstyle 2500\;\sang}$}}
                     \else{$\ell\raise 0.3 em \hbox{$_{\scriptstyle 2500\;\sang}$}$} \fi}
\newcommand{\fuv}   {\ifmmode{f\raise 0.3 em \hbox{$_{\scriptstyle 2500\;\sang}$}}
                     \else{$f\raise 0.3 em \hbox{$_{\scriptstyle 2500\;\sang}$}$} \fi}
\newcommand{\fuvpr} {\ifmmode{f^{\prime}\!\!\raise 0.3 em \hbox{$_{\scriptstyle 2500\;\sang}$}}
                     \else{$f^{\prime}\!\!\raise 0.3 em \hbox{$_{\scriptstyle 2500\;\sang}$}$} \fi}

\newcommand{\fuvintr} {\ifmmode{f^{\;\rm intr}\!\!\!\!\!\!\!\!\raise 0.3 em \hbox{$_{\scriptstyle 2500\;\sang}$}}
                     \else{$f^{\prime}\!\!\!\!\!\!\!\!\raise 0.3 em \hbox{$_{\scriptstyle 2500\;\sang}$}$} \fi}

\newcommand{\fuvobs} {\ifmmode{f^{\;\rm obs}\!\!\!\!\!\!\!\!\raise 0.3 em \hbox{$_{\scriptstyle 2500\;\sang}$}}
                     \else{$f^{\prime}\!\!\!\!\!\!\!\!\raise 0.3 em \hbox{$_{\scriptstyle 2500\;\sang}$}$} \fi}

\newcommand{\fx}    {{\ifmmode{f_{\rm 2\; keV}}             \else{$f_{\rm 2\; keV}$}\fi}}
\newcommand{\fxpr}  {{\ifmmode{f_{\rm 2\; keV}^{\prime}}    \else{$f_{\rm 2\; keV}^{\prime}$}\fi}}

\def\ls{\lower0.3em\hbox{$\,\buildrel <\over\sim\,$}}
\def\gs{\lower0.3em\hbox{$\,\buildrel >\over\sim\,$}}

\newcounter{species} 
\def\ion#1#2{\setcounter{species}{#2}#1$\;${\scriptsize\Roman{species}}\relax}

\def\chandra{{\it Chandra}}
\def\xmm{{\it XMM-Newton}}

\def\skp{\noalign{\vskip 6pt}}

\slugcomment{Accepted for publication in the Astrophysical Journal}
\shorttitle{Exploratory Study of Intrinsic-NAL Quasars}
\shortauthors{Misawa et al.}

\begin{document}

\title{Exploratory Study of the X-Ray Properties of Quasars With
Intrinsic Narrow Absorption Lines}

\author{Toru Misawa,
        Michael Eracleous\altaffilmark{1},
        George Chartas, and
        Jane C. Charlton}

\affil{Department of Astronomy and Astrophysics, Pennsylvania State
  University, 525 Davey Lab, University Park, PA 16802}

\altaffiltext{1}{Center for Gravitational Wave Physics,
  The Pennsylvania State University, University Park, PA 16802}

\email{{\tt misawa,mce,chartas,charlton@astro.psu.edu}}

\begin{abstract}
We have used archival \chandra\ and \xmm\ observations of quasars
hosting intrinsic narrow UV absorption lines (intrinsic NALs) to carry
out an exploratory survey of their X-ray properties. Our sample
consists of three intrinsic-NAL quasars and one ``mini-BAL'' quasar,
plus four quasars without intrinsic absorption lines for comparison.
These were drawn in a systematic manner from an optical/UV-selected
sample. The X-ray properties of intrinsic-NAL quasars are
indistinguishable from those of ``normal'' quasars. We do not find any
excess absorption in quasars with intrinsic NALs, with upper limits of
$\nh \ls{\rm a\; few}\times 10^{22}~\cmm$.  We compare the X-ray and
UV properties of our sample quasars by plotting the equivalent width
and blueshift velocity of the intrinsic NALs and the X-ray spectral
index against the ``optical-to-X-ray'' slope, \aox. When BAL quasars
and other AGNs with intrinsic NALs are included, the plots suggest
that intrinsic-NAL quasars form an extension of the BAL sequences and
tend to bridge the gap between BAL and ``normal''
quasars. Observations of larger samples of intrinsic-NAL quasars are
needed to verify these conclusions. We also test two competing
scenarios for the location of the NAL gas in an accretion-disk
wind. Our results strongly support a location of the NAL gas at high
latitudes above the disk, closer to the disk axis than the dense BAL
wind.  We detect excess X-ray absorption only in Q0014+8118, which
does not host intrinsic NALs.  The absorbing medium very likely
corresponds to an intervening system at $z=1.1$, which also produces
strong absorption lines in the rest-frame UV spectrum of this
quasar. In the appendix we discuss the connection between UV and X-ray
attenuation and its effect on \aox.
\end{abstract}


\section{Introduction}\label{sec:intro}

The {\it intrinsic} absorption lines found in the rest-frame UV
spectra of quasars and active galactic nuclei (AGNs) are classified
based on their line widths into broad absorption lines (BALs;
FWHM$\,>2000$~\kms, found in $\sim 10\%$ of all quasars), narrow
absorption lines (NALs; FWHM$\,< 500$~\kms, found in $\sim 50\%$ of
quasars at $z\sim 2$--4; these can be clearly separated from
intervening lines via high-resolution spectroscopy), and mini-BALs
(intermediate FWHM between BALs and NALs). They are typically
blueshifted relative to the quasar and are thought to trace outflows
from the quasar central engine. These outflows could be hydromagnetic
accretion-disk winds (e.g., Blandford \& Payne 1982; Emmering,
Blandford \& Shlosman 1992; K\"onigl \& Kartje 1994; Everett 2005), or
accretion-disk winds driven by radiation pressure (e.g., Murray et
al. 1995; Arav, Li, \& Begelman 1994, Proga, Stone, \& Kallman 2000).
Alternatively they could be outflows driven by thermal pressure and
launched either from the accretion disk itself (e.g., Begelman, McKee,
\& Shields 1983) or from the obscuring torus invoked by AGN
unification schemes and driven by thermal pressure (e.g., Balsara \&
Krolik 1993; Krolik \& Kriss 1995, 2001; Chelouche \& Netzer
2005). However, thermal accretion disk winds are too hot to produce
absorption lines in the UV, while thermal winds from the obscuring
torus are too slow to account for the observe velocities of NALs and
BALs. Accretion disk winds, whatever their origin, are an integral
part of the accretion flow, and may be the source of the broad
emission lines that are characteristic of the optical and UV spectra
of quasars and AGNs (e.g., Shields 1977; Chiang \& Murray 1996; Murray
\& Chiang 1997). Moreover, the outflows may be important for cosmology
since they deliver energy and momentum to the interstellar and
intergalactic media (hereafter, ISM and IGM, respectively) and can
affect galaxy evolution (e.g., Granato et al. 2004; Scannapieco \& Oh
2004; Springel, Di Matteo \& Hernquist 2005; Chartas et al. 2007a).

Most of our knowledge about intrinsic absorbers is derived from
optical/UV observations. A combination of observational results and
models suggest that intrinsic absorption lines of different widths
represent either different lines of sight through the outflowing wind
to the quasar continuum source (Ganguly et al. 2001; Elvis 2000), or
different stages in the evolution of the absorbing gas parcels (e.g.,
Hamann \& Sabra 2004; Misawa et al. 2005).

X-ray observations of BAL quasars have revealed large columns of
nearly-neutral absorbing gas ($\sim 10^{23-24}$~\cmm; see, for
example, Green \& Mathur 1996; Gallagher et al. 2002a).  However,
little is known about the X-ray properties of quasars with intrinsic
NALs and mini-BALs. X-ray spectroscopy of the last two types of
quasars can be extremely useful in many respects:

\begin{itemize}

\item
In general terms, we can compare the X-ray properties of quasars that
host intrinsic NALs with those that do not. This comparison allows us
to assess whether there is a significant difference between the
central engines of the two types of object. If no difference is found,
and since intrinsic NALs are ubiquitous in quasar spectra (see Misawa
et al 2007b and references therein), we will be led to a picture where
intrinsic NALs, hence outflows, are a universal property of quasars.

\item
X-ray spectra can probe highly-ionized media that are not directly
probed by optical/UV absorption lines. Moreover, they can yield more
direct measurements of the total hydrogen column density of absorbers
in a relatively low-ionization state (this cannot be easily done
through UV absorption lines, unless the ionization state of the gas is
very well constrained). Thus, we can use the X-ray spectra to
investigate whether any hot gas (which may represent the bulk of the
mass) co-exists (mixed or layered) with the UV absorbers.
\item
The column densities determined from the X-ray spectra constitute a
direct test of scenarios for the location of the NAL gas in the larger
outflow. In particular, Elvis (2000) suggests that the NAL gas is
located at very low latitudes above the accretion disk, implying very
large column densities (comparable to those found in BAL quasars,
i.e., $\gs 10^{22}$~\cmm). On the other hand, Ganguly et al. (2001)
place the NAL gas at high latitudes above the accretion disk,
implicitly suggesting lower column densities.

\item
Several examples of intrinsic NALs at high ejection velocities ($\sim
0.2c$) are now known (see Misawa et al. 2007b and references therein).
It is extremely interesting, therefore, to search quasars with
intrinsic NALs for high-velocity X-ray absorption lines (see, for
example, Chartas et al. 2002, 2003). Such observations can yield 
results relevant to cosmology since they can lead to estimates of 
the mass outflow rate and the kinetic power of the outflow (see,
for example, Chartas et al. 2007b).

\end{itemize}

With the above considerations in mind, we have used archival \chandra\
and \xmm\ data to carry out an exploratory survey of the X-ray
properties of a small, but carefully-selected sample of quasars at
$z\approx 2.5$--3.8.  This sample consists of three quasars hosting
intrinsic NALs, one quasar hosting a mini-BAL (HS1603+3820), and four
quasars without intrinsic absorption lines for comparison (these were
selected in a systematic way, as described in the next section). We
compare the properties of the UV NALs (e.g., equivalent width, outflow
velocity) with the parameters describing the X-ray spectrum (e.g.,
photon index, {\it intrinsic} column density, and optical-to-X-ray
slope, \aox).  Thus, we place NAL quasars in the context of BAL
quasars (see Gallagher et al. 2002b; Brandt, Laor, \& Wills 2000) and
investigate whether all types of objects follow the same trends in
their properties. We also use the column densities determined from the
X-ray spectra to carry out one of the tests outlined, above, namely,
we attempt to distinguish between the two different suggestions for
the location of the NAL gas in the larger outflow.

In \S\ref{sec:sample}, we describe how the sample was selected and
summarize the properties of the constituent quasars.  In
\S\ref{sec:obs} and \S\ref{sec:analysis}, we present the analysis of
the data (observations, data screening, and model fits to the X-ray
spectra). We compare the rest-frame UV and X-ray properties in
\S\ref{sec:xuv}.  In \S\ref{sec:discussion} we summarize our findings,
discuss our results, and consider prospects for future work.  We adopt
a cosmological model with $H_{0}$ = 75~\kms~Mpc$^{-1}$,
$\Omega_{m}$=0.3, and $\Omega_{\Lambda}$ = 0.7. Throughout this paper
we give error bars corresponding to the 90\% confidence level, unless
noted otherwise. A significant part of our discussion makes use of the
optical-to-X-ray slope of the spectral energy distribution, \aox,
which is affected by extinction. Thus we examine the effects of
extinction on \aox\ in detail in the appendix.

\section{Sample Selection}\label{sec:sample}

Our sample was drawn from a survey for intrinsic NALs in the Keck
high-resolution spectra of 37 quasars at $z=2$--4 (Misawa et
al. 2007b). These quasars were originally selected and observed in
order to study intergalactic deuterium lines (e.g., O'Meara et
al. 2001, and references therein), therefore, they make up a largely
unbiased sample with respect to {\it intrinsic} NALs. The intrinsic
NALs were separated from NALs arising in cosmologically intervening
objects based on their partial coverage signature (e.g., Hamann et
al. 1997b; Barlow \& Sargent 1997; Ganguly et al. 1999, 2003; Misawa
et al. 2003, 2005). In summary, the partial coverage test exploits the
fact that the ratio of optical depths in the individual lines of UV
resonance doublets is prescribed by atomic physics. Departures from
the expected ratio in well-resolved, unsaturated doublets is
interpreted as the result of dilution of the line troughs by continuum
photons that do not pass through the absorber (e.g., Hamann et
al. 1997b; Barlow \& Sargent 1997; Ganguly et al. 1999), i.e., the
absorber is very compact, thus intrinsic to the quasars (most
cosmologically intervening structures are considerably more extended
than the quasar continuum source). In the case of the prominent UV
resonance doublets (\ion{C}{4}, \ion{Si}{4}, and \ion{N}{5}), the
ratio of optical depths between doublet members is 2:1, which allows
us to compute analytically the fraction of the continuum photons that
pass through the absorber (the coverage fraction). In practice
the coverage fraction is computed both by applying the above algorithm
to every pixel of the line profile and by decomposing the profile into
kinematic components (represented by Voigt functions whose FWHM turn
out to be of order 10~\kms) and applying the above algorithm to each
component. If both methods consistently reveal partial coverage in
parts of the line profile, the line is identified as intrinsic. We
find that the profiles of the \ion{C}{4} absorption lines often have
multiple kinematic components, not sharing the same coverage
fraction. Further details of the methodology can be found in Misawa
et al (2007b), and the references cited therein.

The above search yielded intrinsic \ion{C}{4} NALs in 32\% of the
quasars and intrinsic \ion{C}{4}, \ion{Si}{4}, or \ion{N}{5} NALs in
50\% of the quasars.  To this sample we added one more quasar hosting
a \ion{C}{4} mini-BAL showing the signature partial coverage as well
as variability, HS1603+3820 (Misawa et al. 2007a, and references
therein). After searching the \chandra\ and \xmm\ archives we found
that 11 quasars from the above survey had been observed and 8 of the
observed quasars had archival X-ray spectra with enough counts to
yield interesting spectral constraints (the 3 objects that were
excluded from our final sample are: Q0241$-$0146, Q0636$+$6801, and
Q1055$+$4611). The 8 quasars making up the final sample are listed in
Table~\ref{tab:quasars}, which also summarizes their basic properties
and the properties of their intrinsic NALs. Of these, three objects
have intrinsic NALs, one has a mini-BAL, and four have no intrinsic
absorption lines.   Thus we have a small, but well defined sample of NAL
quasars and a matching comparison sample of quasars without intrinsic
NALs.  All of the quasars with intrinsic NALs and 2/4 quasars without
intrinsic NALs are radio quiet.  The \chandra\ and \xmm\ X-ray spectra
of 6 of these quasars were analyzed and the results published (see
references in \S\ref{sec:notes}). Nevertheless, we have re-analyzed
the data for the entire sample for the sake of uniformity of analysis
and in order to make a careful assessment of absorption intrinsic to
the quasar central engines. To this end we have tested a variety of
spectral models, as we detail in \S\ref{sec:models}.

\begin{deluxetable*}{lcccccrcl}[b]
\tabletypesize{\scriptsize}
\tablecaption{Sample Quasars \label{tab:quasars}}
\tablewidth{0pt}
\tablehead{
\colhead{} & 
\multicolumn{4}{c}{Quasar Properties} & 
\multicolumn{4}{c}{Intrinsic \ion{C}{4} NAL Properties} \\
\colhead{} & 
\multicolumn{4}{c}{\hrulefill} & 
\multicolumn{4}{c}{\hrulefill} \\
\colhead{} & 
\colhead{} & 
\colhead{Radio} & 
\colhead{$D_{\rm L}$\tablenotemark{b}} &
\colhead{\nhgal} & 
\colhead{\fuv\tablenotemark{c}} &
\colhead{\voff\tablenotemark{d}} & 
\colhead{Rest EW} &
\colhead{\cf\tablenotemark{e}} \\
\colhead{Quasar} & 
\colhead{$z$} & 
\colhead{Loud?\tablenotemark{a}} & 
\colhead{(Gpc)} &
\colhead{($10^{20}$~\cmm)} & 
\colhead{(mJy)} &
\colhead{(\kms)} & 
\colhead{(\AA)} &
\colhead{} \\
\colhead{(1)} & 
\colhead{(2)} & 
\colhead{(3)} &
\colhead{(4)} &
\colhead{(5)} & 
\colhead{(6)} &
\colhead{(7)} &
\colhead{(8)} &
\colhead{(9)} 
}
\startdata
Q0014$+$8118  & 3.387 & L & 27.4 &  14 & 0.70 &  \dots    & \dots                       & \dots      \\
Q0130$-$4021  & 3.030 & Q & 24.0 & 1.9 & 0.21 & $-65,181$ & 0.53                        & $(0.19\pm0.03)$, $(0.83^{+0.10}_{-0.09})$, $(0.83^{+0.16}_{-0.15})$, 1 \\
              &       &   &      &     &      & $-37,035$ & 0.69\tablenotemark{f}       & 1, 1, 1, $(0.67\pm0.03)$, $(0.6\pm0.1)$, $(0.27\pm0.03)$, 1 \\
Q1107$+$4847  & 3.000 & Q & 23.7 & 1.3 & 0.30 & $-21,388$ & 0.20                        & 1, $(0.70^{+0.17}_{-0.15})$, 1 \\
Q1208$+$1011  & 3.803 & Q & 31.5 & 1.7 & 0.04 &  \dots    & \dots                       & \dots      \\
Q1422$+$2309  & 3.611 & L & 29.6 & 2.7 & 0.05 &  \dots    & \dots                       & \dots      \\
Q1442$+$2931  & 2.670 & Q & 20.6 & 1.7 & 0.47 &  \dots    & \dots                       & \dots      \\
HS1603$+$3820\tablenotemark{g} & 2.542 & Q & 19.4 & 1.3 & 0.77 & $-32,657$ & 0.33       & 1, 1, 1, $(0.70^{+0.24}_{-0.16})$, 1, $(0.87\pm0.02)$, 1 \\
              &       &   &      &     &      & $ -9,000$ & 2.53--5.07\tablenotemark{g} & 0.24--0.45 ($\pm0.02$)\tablenotemark{g}\\
              &       &   &      &     &      &    $+800$ & 0.74                        & $(0.30^{+0.14}_{-0.09})$, 1,  $(0.23\pm0.05)$, 1, 1 \\
Q1700$+$6416  & 2.722 & Q &  21.1& 2.7 & 0.53 & $-24,195$ & 0.58\tablenotemark{h}       & 1.00       \\
              &       &   &      &     &      & $-23,641$ &  ...\tablenotemark{h}       & 1.00       \\
              &       &   &      &     &      &    $-765$ & 0.14                        & $(0.49\pm0.06)$, 1 \\
\enddata
\tablenotetext{a}{Indicates whether the quasar is radio-loud (L) or
                  radio-quiet (Q); from Misawa et al. (2007b).}
\tablenotetext{b}{The luminosity distance, computed based on the
                  cosmological parameters given at the end of
                  \S\ref{sec:intro} of the text.}
\tablenotetext{c}{Flux density per unit frequency at 2500~\AA\ in the
                  quasar rest frame, derived from the V- or R-band
                  flux (assuming $f_{\nu}\propto \nu^{-0.44}$; see
                  Misawa et al. 2007b) and corrected for Galactic
                  extinction but not intrinsic extinction.}
\tablenotetext{d}{Velocity offset of a NAL relative to the redshift of
                  the quasar. A negative value denotes a blueshift.}
\tablenotetext{e}{Coverage fractions for the kinematic (Voigt)
                  components used to decompose the line profile. Each
                  component has a FWHM of order 10~\kms (from Misawa
                  et al. 2007a,b; see \S2 of the text). These values
                  effective represent the variation of \cf\ across the
                  \ion{C}{4} line profile.}
\tablenotetext{f}{The red member of this doublet is blended with
                  another line, but the equivalent width can be
                  evaluated.}
\tablenotetext{g}{This quasar hosts a \ion{C}{4} mini-BAL at a
                  blueshifted velocity of $-9,000~\kms$. The
                  rest-frame equivalent width of this mini-BAL varies
                  significantly over short time scales (see Misawa et
                  al. 2007a). Thus we list a range of \cf\ values to 
                  indicate its temporal variability (the error bars 
                  are comparable over all epochs, hence we give only a
                  representative value here).}
\tablenotetext{h}{These lines are locked to each other and as
                  such they are judged to be intrinsic. The coverage
                  fraction is unity and only the total equivalent
                  width can be evaluated.}
\end{deluxetable*}

In Table~\ref{tab:quasars} we give the quasar name, redshift, Galactic
column density (\nhgal), and flux density at 2500~\AA\ in the quasar
rest frame, after correcting for absorption in the ISM of the Milky
Way, but without any correction for intrinsic absorption.  We also
include the properties of intrinsic NALs, namely the offset velocity
relative to the quasar redshift\footnote{Absorption-line systems close
to the systemic redshift of the quasar are regarded as ``associated''
with the quasar, just on the basis of their proximity in velocity
space. However, different authors adopt a different velocity limit for
their definition of associated systems. The most common convention is
$|\voff| \leq 5,000$~\kms\ (e.g., Foltz et al 1986), while Brandt et
al. (2000) adopt $|\voff| \leq 12,000$~\kms.}, the rest-frame
equivalent width of the intrinsic \ion{C}{4} NALs, and the coverage
fraction of the absorber. The coverage fraction was determined
as described in the first paragraph of this section. In
Table~\ref{tab:quasars} we list the coverage fractions of the Voigt
components used to decompose each \ion{C}{4} line profile.  These
values effectively represent the variation of the coverage fraction
across the \ion{C}{4} profiles in bins of width of order 10~\kms and
provide the basis for the classification of the NALs as intrinsic 
(the coverage fraction is significantly lower than unity in large
portions of the line profile, especially close to the line center).
The classification of these NALs as intrinsic is also corroborated by
similar partial coverage information on other lines in the same
system, such as \ion{N}{5}. In the case of Q1700+6416, the
higher-velocity absorption system listed in Table~\ref{tab:quasars} is
judged to be intrinsic based on the fact that it is line locked.

\section{X-Ray Observations and Data Screening}\label{sec:obs}

The 8 quasars in our sample were observed with the \chandra\ Advanced
CCD Imaging Spectrometer (ACIS; Garmire et al. 2003) and/or with the
\xmm\ European Photon Imaging Camera (EPIC) pn and MOS detectors
(Str\"uder et al. 2001; Turner et al. 2001). Some objects were
observed with both instruments and some objects were observed more
than once with the same instrument, yielding a total of 6 \chandra\
and 6 \xmm\ data sets.  In Table~\ref{tab:obs} we give a log of the
observations, including the observatory and instrument, the
observation identification number, and the date.

\begin{deluxetable*}{lllclcr}
\tabletypesize{\scriptsize}
\tablecaption{Observation Log\label{tab:obs}}
\tablewidth{0pt}
\tablehead{
\colhead{} &
\colhead{} &
\colhead{} &
\colhead{} &
\colhead{} & 
\colhead{Exp.} & 
\colhead{} \\
\colhead{Quasar} &
\colhead{Observatory} &
\colhead{Instr.} &
\colhead{Obs.ID} &
\colhead{Date} & 
\colhead{(ks)} &
\colhead{Counts} \\
\colhead{(1)} &
\colhead{(2)} &
\colhead{(3)} &
\colhead{(4)} &
\colhead{(5)} &
\colhead{(6)} &
\colhead{(7)} 
}
\startdata
Q0014+8118  & \xmm\    & pn        & 0112620201 & 2001 Aug 23 & 14.3 & 13348  \\
            &          & MOS1      &            &             & 16.9 &  4866  \\
            &          & MOS2      &            &             & 16.9 &  4827  \\
Q0130-4021  & \xmm\    & pn        & 0112630201 & 2001 Jun 04 & 23.0 &  1842  \\
            &          & MOS1      &            &             & 25.2 &   499  \\
            &          & MOS2      &            &             & 25.3 &   491  \\
Q1107+4847  & \xmm\    & pn        & 0104861001 & 2002 Jun 01 & 28.1 &   813  \\
            &          & MOS1      &            &             & 32.2 &   213  \\
            &          & MOS2      &            &             & 32.3 &   253  \\
Q1208+1011  & \chandra & ACIS-S    & 3570       & 2003 Mar 02 & 9.97 &   175  \\
Q1422+2309  & \chandra & ACIS-S    & 367        & 2000 Jun 01 & 28.4 &   426  \\
            & \chandra & ACIS-S    & 1631       & 2001 May 21 & 10.7 &   244  \\
            & \xmm\    & pn        & 0143652301 & 2003 Feb 04 &  ...\tablenotemark{a} &   ...\tablenotemark{a} \\
            &          & MOS1      &            &             & 4.89 &   191  \\
            &          & MOS2      &            &             & 4.89 &   198  \\
            & \chandra & ACIS-S    & 4939       & 2004 Dec 01 & 47.7 &   762  \\
Q1442+2931  & \xmm\    & pn        & 0103060201 & 2002 Aug 01 & 21.3 &  2607  \\
            &          & MOS1      &            &             & 24.8 &   768  \\
            &          & MOS2      &            &             & 24.8 &   753  \\
HS1603+3820 & \chandra & ACIS-S    & 4026       & 2002 Nov 29 & 8.30 &   137  \\
Q1700+6416  & \chandra & ACIS-S    & 547        & 2000 Oct 31 & 49.5 &   325  \\
            & \xmm\    & pn        & 0107860301 & 2002 May 31 & 14.1 &  1060  \\
            &          & MOS1      &            &             & 18.3 &   281  \\
            &          & MOS2      &            &             & 18.4 &   299  \\
\enddata
\tablenotetext{a}{The pn image of Q1422+2309 fell on a chip gap,
                    therefore we were not able to use those data.}
\end{deluxetable*}

We retrieved the data from the respective data archives and reduced
them in a standard and uniform manner using up-to-date calibration
data\footnote {Updates on the calibration of Chandra and XMM-Newton
are available on the WWW sites of the CXC ({\tt
http://asc.harvard.edu/ciao/releasenotes/history.html}) and \xmm\ SOC
({\tt
http://xmm.vilspa.esa.es/external/ xmm\_sw\_cal/calib/rel\_notes/index.shtml})}.
The \chandra\ data were reduced using the CIAO~3.3 software package,
provided by the \chandra\ X-ray Center (CXC) and following the CXC
threads.  We screened the data according to status, photon event
grade, aspect solution, and background level. We also removed the
$\pm$0{\farcs}25 randomization applied to the photon positions by the
CXC in order to improve the spatial resolution. The \xmm\ data were
reduced with the SAS~6.5 software package provided by the \xmm\
Science Operation Center (SOC). We screened the pn and MOS data to
retain events with values of the {\tt PATTERN} keyword in the range
0--4 and 0--12, respectively.  To avoid intervals of high background
we excluded data taken when the full-field count rates exceeded 20
s$^{-1}$ in the pn and and 4 s$^{-1}$ in the MOS.

We extracted spectra of the target quasars after subtracting the local
background. In Table~\ref{tab:obs}, we list the exposure times and
number of the counts in the spectra of individual instruments. With
the exception of the \chandra\ spectra of Q1422$+$2309 (taken at
three different epochs) there was no evidence of pile up in the
data. Our sample includes two gravitationally lensed quasars,
Q1208$+$1011 and Q1422$+$2309, which required special treatment. We
give the details of how we treated these two objects in
\S\ref{sec:notes}, below. The luminosities we report for these two
quasars have been scaled down according to the magnification factors
reported in \S\ref{sec:notes}.

\section{Model Fits to the Observed Spectra and Results}\label{sec:analysis}

\subsection{Suite of Models and Fit Results}\label{sec:models}

We fitted a variety of models to the observed X-ray spectra using the
XSPEC~12.3.0 software package (Arnaud 1996). All energy channels below
0.4~keV were ignored.  The spectra were binned to have a minimum of 10
counts per bin (in most cases the minimum number of counts per bin was
required to be higher; see \S\ref{sec:notes} for specific details). In
the case of \xmm\ data, we performed joint fits to the MOS and pn
spectra.

We used the suite of models, described below, to fit the spectra and
selected the model that gave the best fit. It is important to find the
most appropriate models for the observed spectra so that we can
determine the intrinsic column density reliably, since this is a
crucial quantity in our subsequent discussion.  Absorption in the
ISM of the Milky Way is included in all models, using the Galactic
column densities from the HEASARC {\tt nH} tool\footnote{\tt
http://heasarc.gsfc.nasa.gov/cgi-bin/Tools/w3nh/w3nh.pl} (\nhgal,
listed in Table~\ref{tab:quasars}) and the photoelectric absorption
cross-sections of Morrison \& McCammon (1983). These column densities
were held fixed during the fitting process. Wherever necessary, we
adopted a Solar abundance pattern for heavy elements from Anders \&
Grevesse (1989).

The models we used are outlined below (along with their XSPEC model
syntax; Galactic absorption is denoted by {\tt wabs}).

\begin{deluxetable*}{rlllclclclcc}
\tabletypesize{\scriptsize}
\tablecaption{Parameters of Best-Fit Models\label{tab:fits}}
\tablewidth{0pt}
\tablehead{
\multicolumn{1}{c}{} &
\multicolumn{2}{c}{Model~\ref{mod:pow}:} &
\multicolumn{3}{c}{Model~\ref{mod:abspow}:} &
\multicolumn{6}{c}{} \\
\multicolumn{1}{c}{} &
\multicolumn{2}{c}{Simple Power Law} &
\multicolumn{3}{c}{Power Law \& Intrinsic Absorption} &
\multicolumn{6}{c}{Complex Models} \\
\multicolumn{1}{c}{} &
\multicolumn{2}{c}{\hrulefill} &
\multicolumn{3}{c}{\hrulefill} &
\multicolumn{6}{c}{\hrulefill} \\
\colhead{} & 
\colhead{} &
\colhead{} &
\colhead{} &
\colhead{$\nh$} &
\colhead{} &
\colhead{} &
\colhead{} &
\colhead{$\nh$} &
\colhead{} &
\colhead{F-Test\tablenotemark{d}} &
\colhead{F-Test\tablenotemark{d}} \\
\colhead{Quasar\tablenotemark{a}} & 
\colhead{$\Gamma$} &
\colhead{$\chi_{\nu}^{2}/\nu$ \tablenotemark{b}} &
\colhead{$\Gamma$} &
\colhead{$(10^{22}\; {\rm cm^{-2})}$} &
\colhead{$\chi_{\nu}^{2}/\nu$ \tablenotemark{b}} &
\colhead{Model\tablenotemark{c}} &
\colhead{$\Gamma$} &
\colhead{$(10^{22}\;{\rm cm^{-2})}$} &
\colhead{$\chi_{\nu}^{2}/\nu$ \tablenotemark{b}} &
\colhead{$P_{\rm F}(1)$} &
\colhead{$P_{\rm F}(2)$} \\
\colhead{(1)} & 
\colhead{(2)} & 
\colhead{(3)} &
\colhead{(4)} &
\colhead{(5)} & 
\colhead{(6)} & 
\colhead{(7)} &
\colhead{(8)} &
\colhead{(9)} &
\colhead{(10)}&
\colhead{(11)}&
\colhead{(12)} \\
}
\startdata
Q0014$+$8118  & $1.43\pm 0.02$         & 1.181/383 & $1.48\pm 0.02$         & $1.1\pm0.3$ & 1.104/380 & 3     & $1.49^{+0.03}_{-0.02}$ &  $1.3^{+4.9}_{-0.2}$ & 1.108/381 & $2\times10^{-6}$ & 0.92 \\
\skp
              &                        &           &                        &                               &          & 4     & $1.48\pm 0.03$      & $1.5^{+3.4}_{-0.6}$ & 1.103/379 & $1\times10^{-6}$ & 0.27 \\
\skp
Q0130$-$4021  & $2.01\pm 0.07$         & 1.033/56  & $1.98^{+0.13}_{-0.06}$ & $< 0.26$       & 1.066/55   & 5     & $2.6^{+0.5}_{-0.6}$ & \dots               & 0.866/55   & 0.0011 & \dots\tablenotemark{e} \\
\skp
Q1107$+$4847  & $1.9\pm 0.1$           & 1.315/44  & $2.0^{+0.3}_{-0.2}$    & $< 1.6$       & 1.321/43   & 6 &   $3.3\pm 0.4$        & $< 1.7$             & 1.157/42   & 0.026 & 0.011 \\
\skp
Q1208$+$1011  & $2.2\pm 0.3$           & 1.106/12  & $2.0^{+0.8}_{-0.2}$    & $< 3.3$       & 1.299/11   & \dots & \dots               & \dots               & \dots  \dots & \dots   \\
\skp
Q1422$+$2309  & $1.55^{+0.15}_{-0.14}$ & 0.543/13  & $1.5^{+0.3}_{-0.2}$    & $< 2.2$       & 0.594/12   & \dots & \dots               & \dots               & \dots & \dots & \dots  \\
\skp
              & $1.7^{+0.3}_{-0.2}$    & 1.175/11  & $1.5^{+0.7}_{-0.2}$    & $< 4.3$       & 1.475/10   & \dots & \dots               & \dots               & \dots & \dots & \dots  \\
\skp
              & $1.4\pm 0.2$           & 0.838/45  & $1.5^{+0.4}_{-0.3}$    & $< 4.5$       & 0.852/44   & \dots & \dots               & \dots               & \dots & \dots & \dots  \\
\skp
              & $1.6\pm 0.1$           & 1.272/11  & $1.6^{+0.3}_{-0.2}$    & $< 2.7$       & 1.430/10   & \dots & \dots               & \dots               & \dots & \dots & \dots  \\
\skp
Q1442$+$2931  & $1.87\pm 0.05$         & 0.738/67  & $1.87^{+0.10}_{-0.06}$ & $< 0.36$       & 0.750/66   & \dots & \dots               & \dots               & \dots & \dots & \dots  \\
\skp
HS1603$+$3820 & $1.9\pm 0.3$           & 0.891/9 & $2.00^{+1.4}_{-0.5}$     & $< 8.4$       & 0.972/8    & \dots & \dots               & \dots               & \dots & \dots & \dots  \\
\skp
Q1700$+$6416  & $2.2\pm 0.2$           & 1.297/12  & $2.3^{+0.5}_{-0.3}$    & $< 5.3$       & 1.407/11   & \dots & \dots               & \dots               & \dots & \dots & \dots  \\
\skp
              & $2.1\pm 0.2$          & 0.951/87  & $2.0^{+0.3}_{-0.2}$    & $< 0.55$       & 0.967/86   & \dots & \dots               & \dots               & \dots & \dots & \dots  \\
\enddata
\tablenotetext{a}{Quasar names and observations are in the same order
                    as in Table~\ref{tab:obs}.}
v%
\tablenotetext{b}{Reduced $\chi^2$ and number of degrees of freedom.}
\tablenotetext{c}{Models are described in detail in \S4 of the text.}
\tablenotetext{d}{F-test probability that more complex models provide
                    a better fit than models 1 and 2, respectively.}
\tablenotetext{e}{Models~\ref{mod:abspow} and \ref{mod:compt} have the
                    same number of free parameters, thus the F-test
                    does not give a meaningful result.  For reference,
                    we note that the values of $\chi^2$ for
                    models~\ref{mod:abspow} and \ref{mod:compt} yield
                    chance probabilities of 0.66 and 0.25,
                    respectively.}
\end{deluxetable*}

\begin{enumerate}

\item
{[\verb=wabs(zpow)=] Simple power-law continuum (at the redshift of
the source) {\it without} intrinsic absorption {\it at the
source}. The free parameters of this model are the photon index,
$\Gamma$, and the normalization (i.e., the photon flux per unit energy
at 1~keV).\label{mod:pow}}

\item 
{[\verb=wabs*zwabs(zpow)=] Simple power-law continuum (at the redshift
of the source) {\it with} intrinsic absorption. This model has an
additional free parameter compared to the previous model, the column
density of the intrinsic absorber, \nh. Simple models of this type can
underestimate the column density of the absorber, if the absorber is
ionized or if it covers the source only partly. Therefore, we also
test the more complex absorption models listed below.\label{mod:abspow}}

\item{[\verb=wabs(zpow)+const*wabs*zwabs(zpow)=] Power-law continuum
(at the redshift of the source) with partial coverage intrinsic
absorption. In this model, the absorber is assumed to cover the
projected area of the X-ray source only partially. The free parameters
are those of the previous model plus the coverage fraction (the
fraction of photons that pass through the
absorber).\label{mod:pcovpow}}

\item{[\verb=wabs*absori(zpow)=] Power-law continuum (at the redshift
of the source) with {\it ionized} intrinsic absorption (following Done
et al. 1992; see also Zdziarski et al. 1995). The free parameters are
the normalization and photon index of the power-law continuum, and the
column density and ionization parameter of the absorber,
$\xi$\,\footnote{$\xi\equiv L/nr^2$, where $L$ is the luminosity
of the source between 5~eV and 300~keV, $r$ is its distance from the
absorber, and $n$ is the hydrogen number density of the absorber}. The
temperature of the absorber was held fixed at 30,000~K, while the iron
abundance was held fixed at the Solar value. \label{mod:absori}}

\item
{[\verb=wabs(pexrav)=] Power-law continuum (with an exponential cutoff
at 100~keV) ``reflected'' from a nearly-neutral slab (with solar
abundances; see George \& Fabian 1991; Magdziarz \& Zdziarski
1995). \label{mod:compt}}

\item
{[\verb=wabs*zwabs(pexrav)=] Power-law continuum ``reflected'' from a
nearly-neutral slab with intrinsic absorption included. \label{mod:abscompt}}

\end{enumerate}

\noindent
In addition to the above models we also experimented with adding a
Gaussian Fe~K$\alpha$ line to the simple power law model as well with
broken power-law models with and without absorption.

\begin{deluxetable}{rlcl}
\tabletypesize{\scriptsize}
\tablecaption{Best-Fit Parameters of Model~\ref{mod:absori} (Ionized Absorber)\tablenotemark{a}\label{tab:iabs}}
\tablewidth{0pt}
\tablehead{
\colhead{} & 
\colhead{} & 
\colhead{$\nh$} &
\colhead{} \\
\colhead{Quasar} & 
\colhead{$\Gamma$} &
\colhead{($10^{22}\; {\rm cm^{-2}}$)} &
\colhead{$\chi^{2}_{\nu} / \nu$\tablenotemark{b}} \\
\colhead{(1)} & 
\colhead{(2)} & 
\colhead{(3)} &
\colhead{(4)} 
}
\startdata
Q0014$+$8118  & $1.48\pm 0.03$         & $1.5^{+3.4}_{-0.6}$ & 1.103/379 \\
\skp
Q0130$-$4021  & $2.00^{+0.15}_{-0.08}$ & $< 7.4$             & 1.000/58  \\
\skp
Q1107$+$4847  & $2.1^{+0.3}_{-0.2}$    & $< 7.8$             & 1.202/46  \\
\skp
Q1442$+$2931  & $1.87^{+0.12}_{-0.06}$ & $< 9.9$             & 0.716/69  \\
\skp
Q1700$+$6416  & $2.0^{+0.4}_{-0.2}$    & $< 13$              & 0.938/89  \\
\enddata
\tablenotetext{a}{This model was fitted only to the \xmm\ spectra
                  because these happen to be the ones with the largest
                  number of counts for each object. The quasar
                  Q1422+2309 is an exception; it is so faint that its
                  \xmm\ spectra did not warrant a fit with this
                  model.}
\tablenotetext{b}{Reduced $\chi^2$ and number of degrees of freedom.}
\end{deluxetable}


\begin{figure*}
\centerline{
  \hfill
  \includegraphics[angle=270,width=6.5cm]{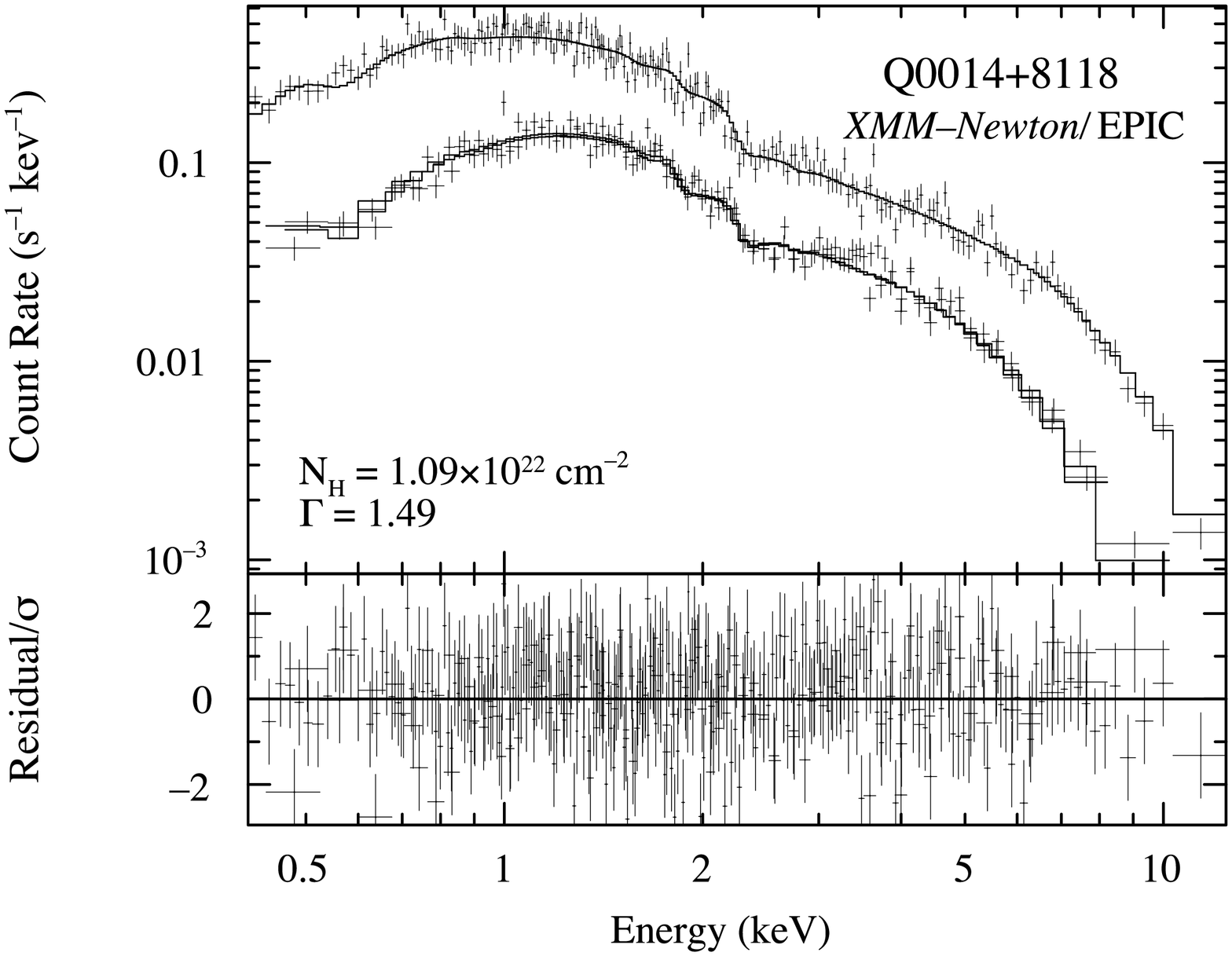}
  \hskip 0.5cm
  \includegraphics[angle=0,width=5.5cm]{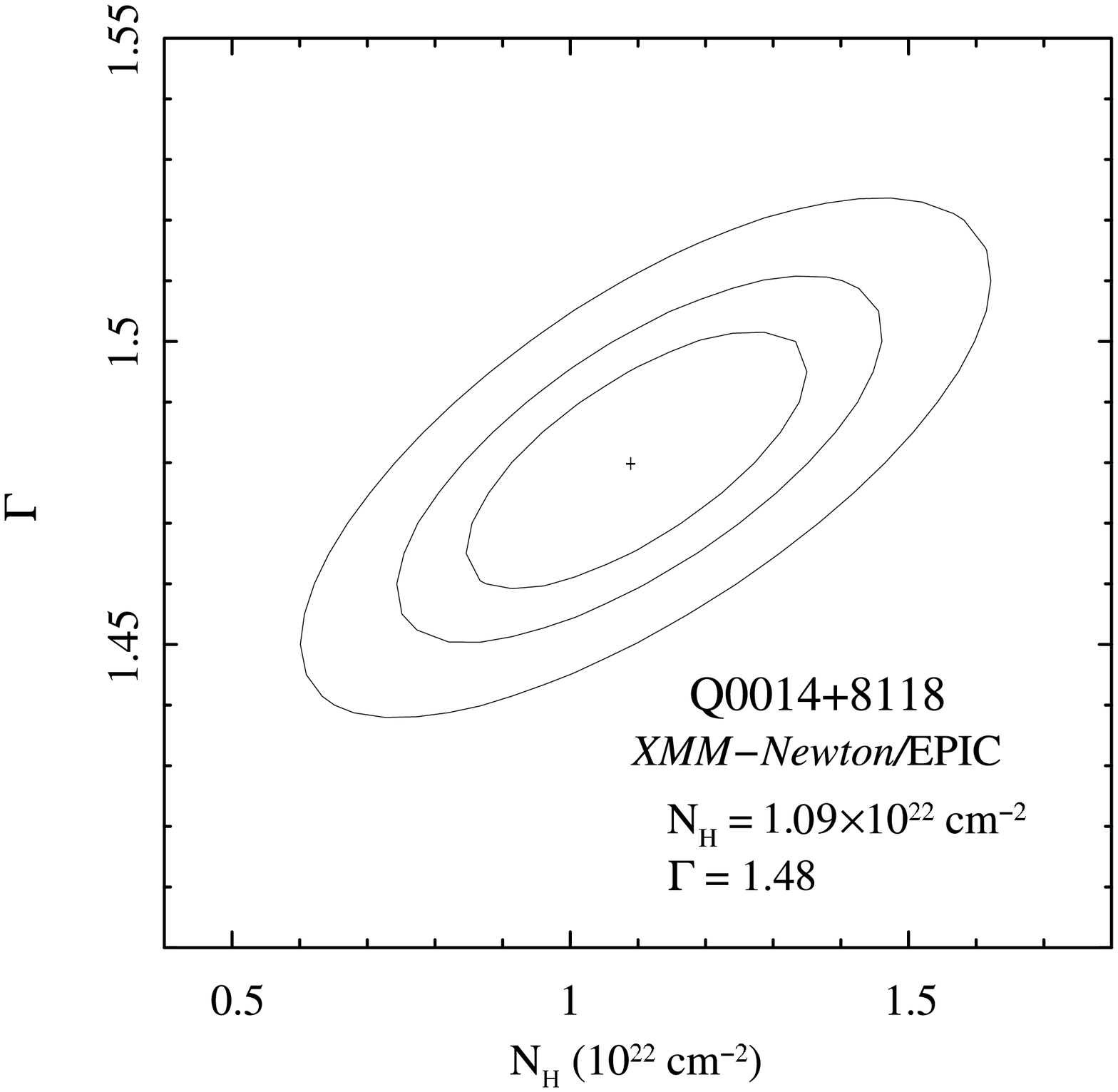}
  \hfill
}
\bigskip
\centerline{
  \hfill
  \includegraphics[angle=270,width=6.5cm]{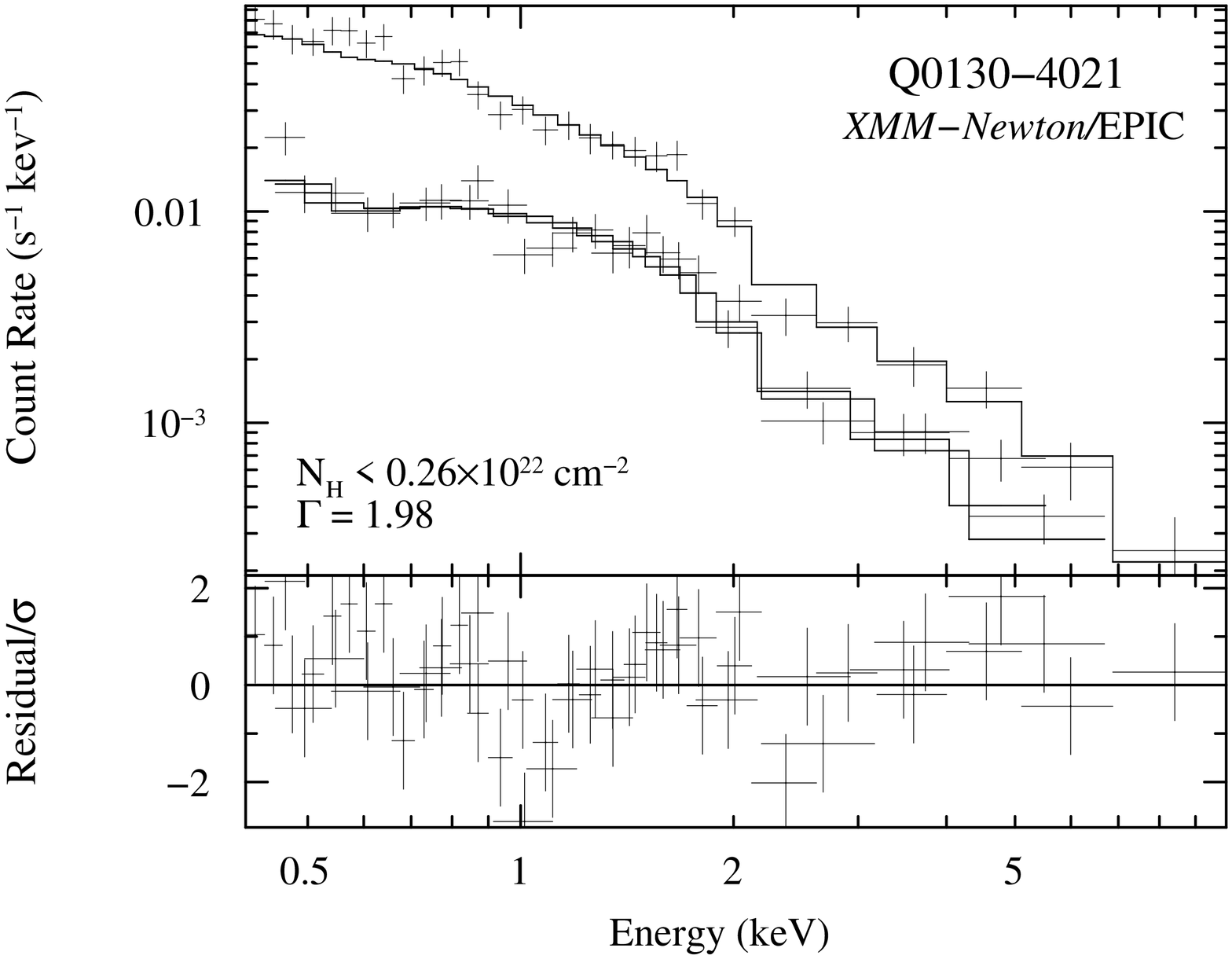}
  \hskip 0.5cm
  \includegraphics[angle=270,width=5.5cm]{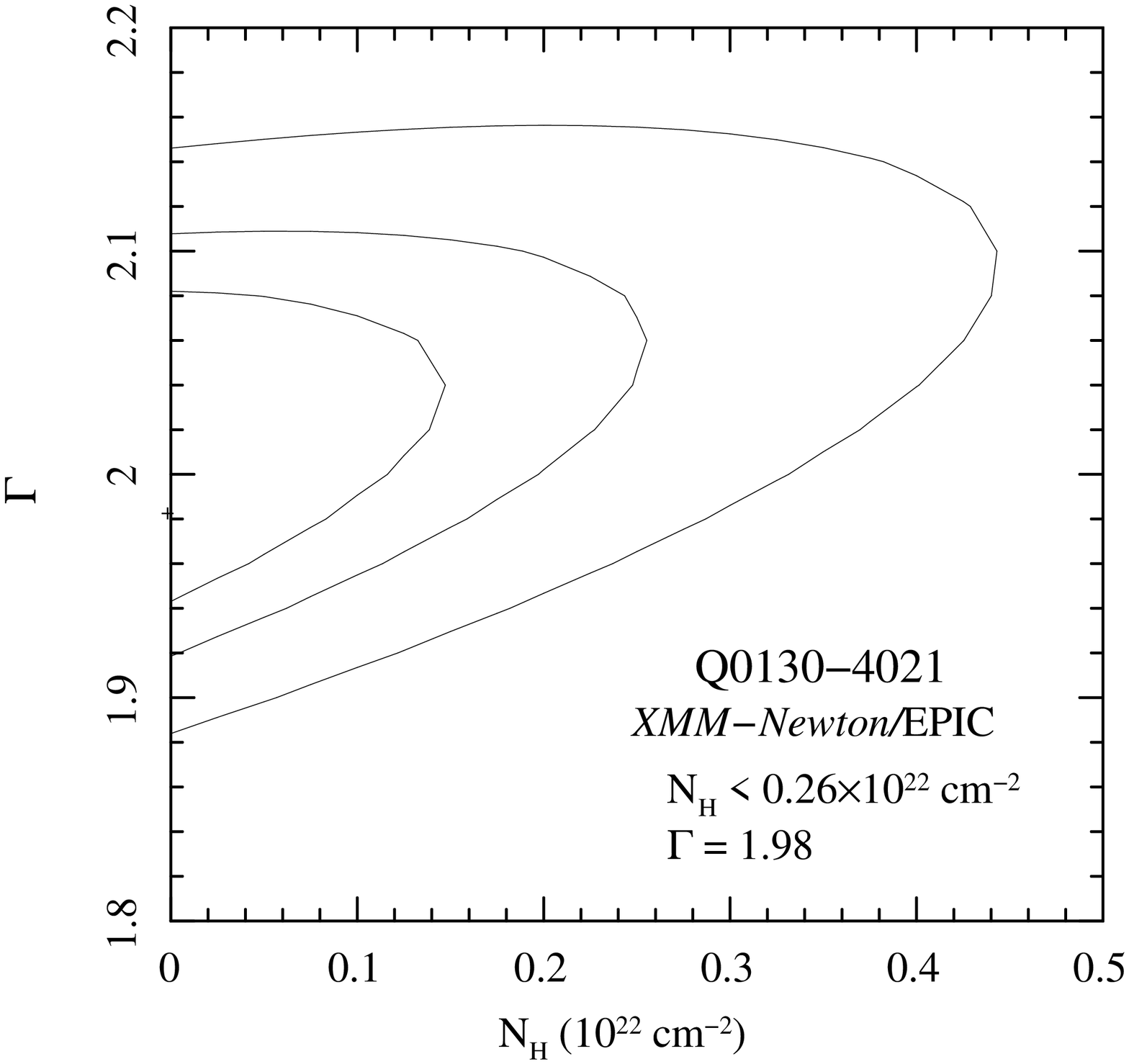}
  \hfill
}
\bigskip
\centerline{
  \hfill
  \includegraphics[angle=270,width=6.5cm]{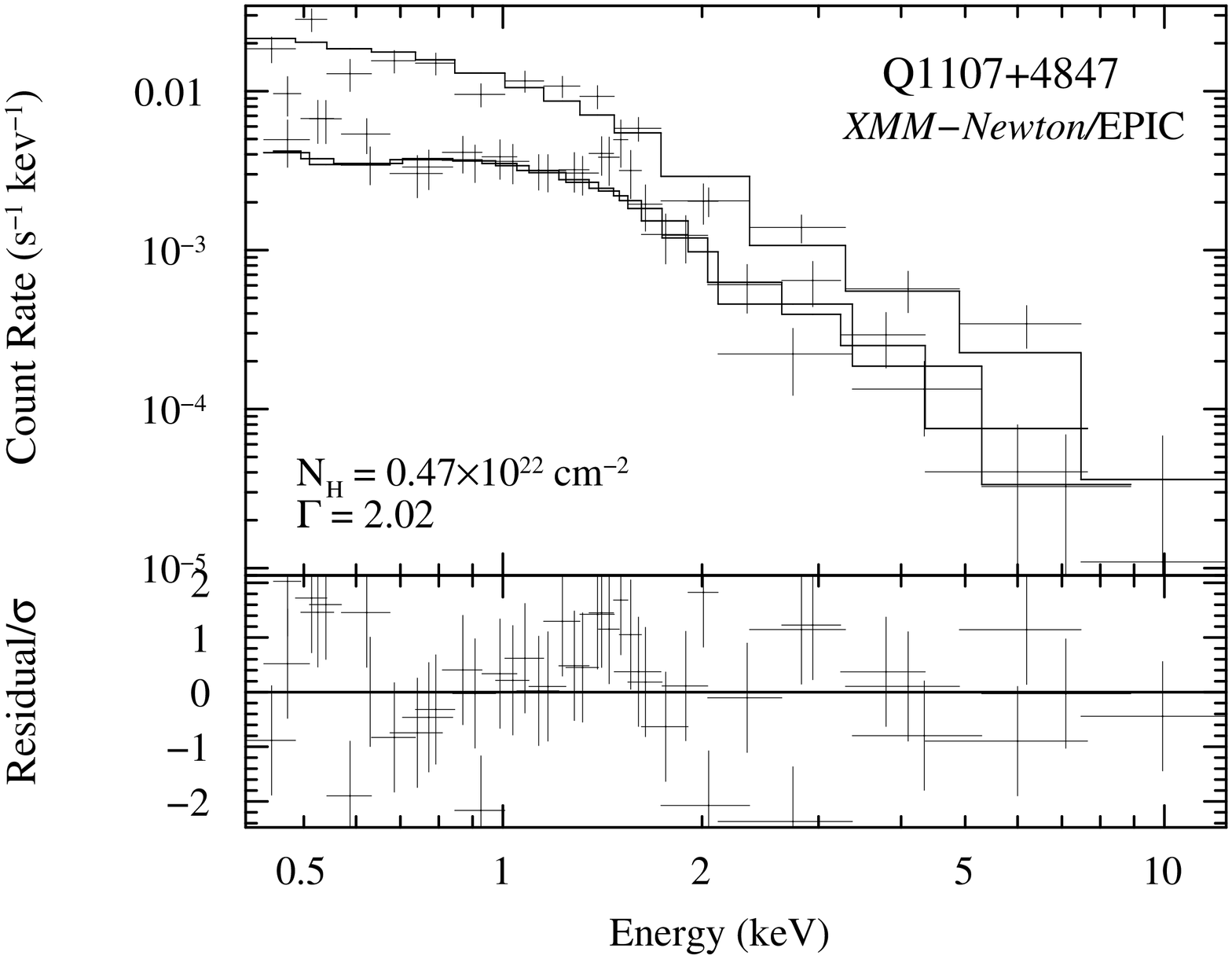}
  \hskip 0.5cm
  \includegraphics[angle=270,width=5.5cm]{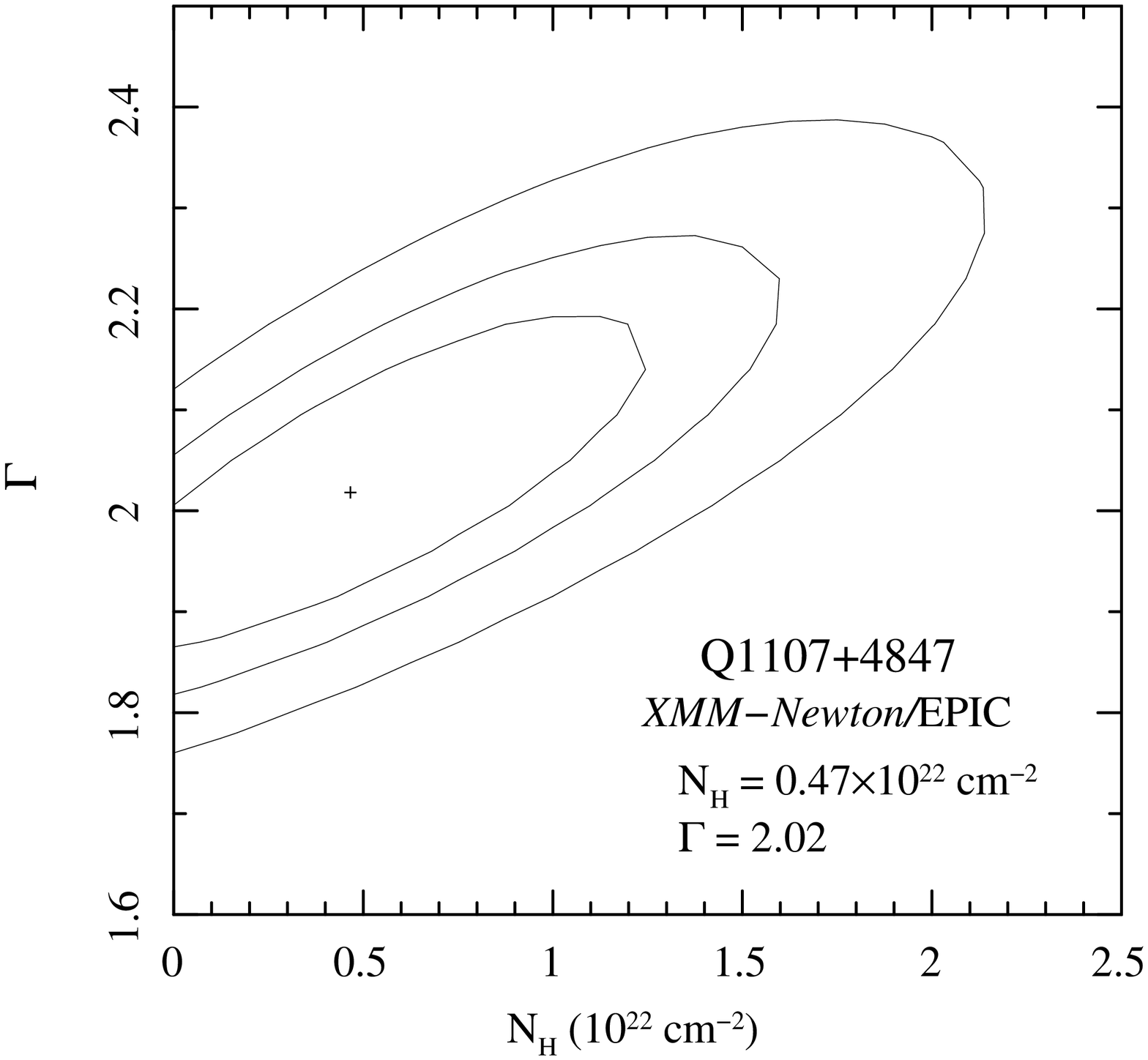}
  \hfill
}
\bigskip
\centerline{
  \hfill
  \includegraphics[angle=270,width=6.5cm]{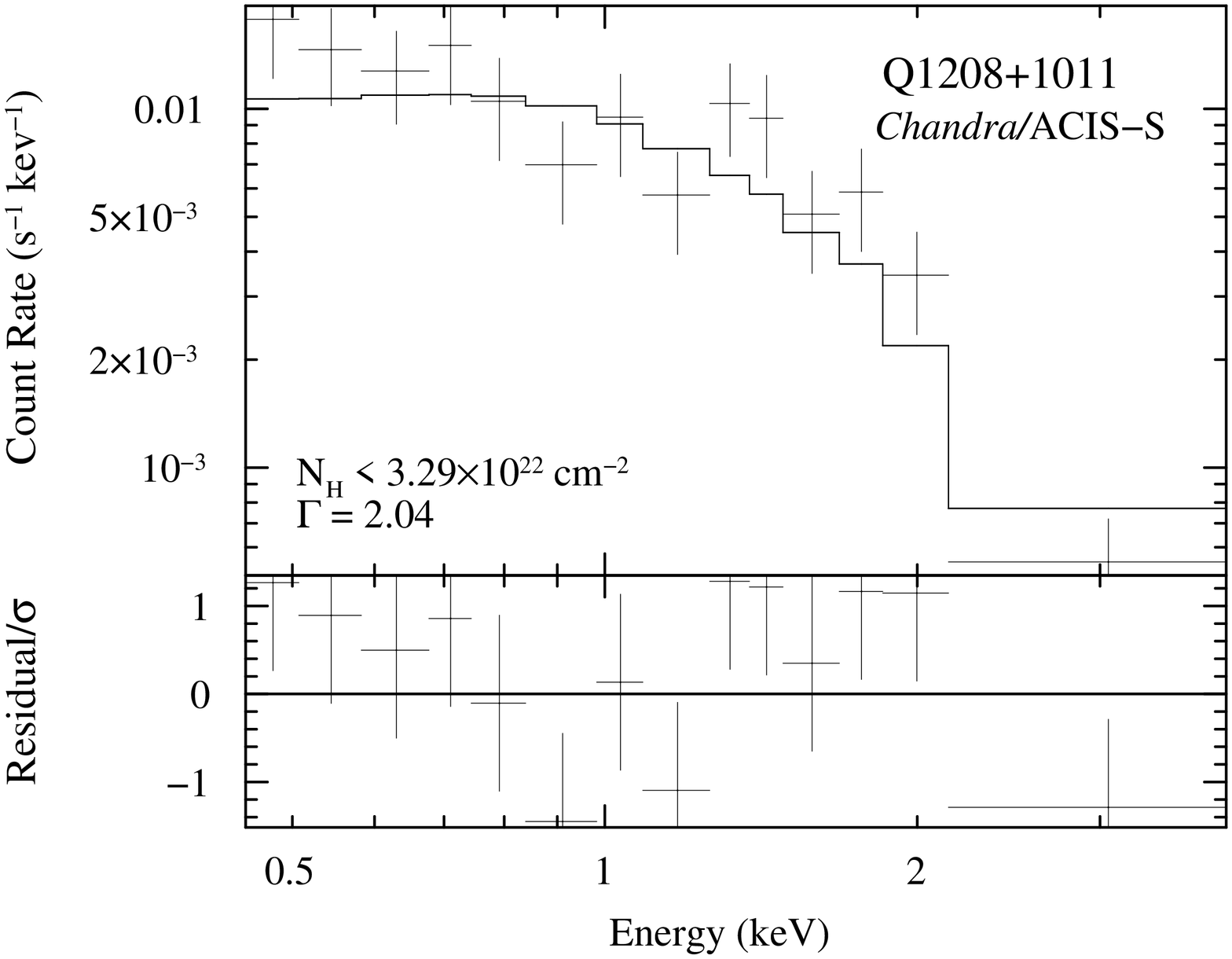}
  \hskip 0.5cm
  \includegraphics[angle=270,width=5.5cm]{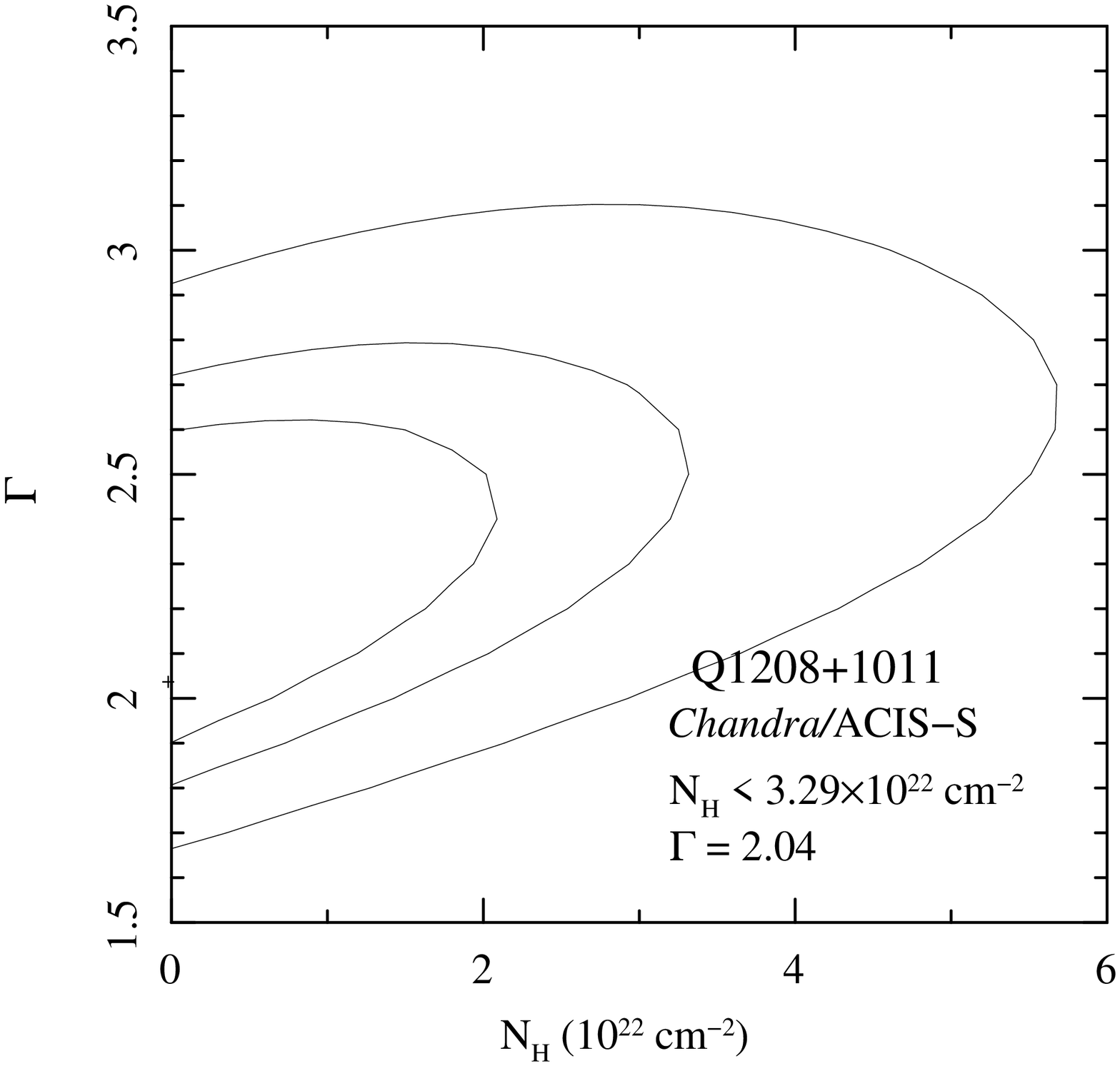}
  \hfill
}
\caption{{\it Left Column:} Observed-frame spectra with the
  best-fitting version of model~\ref{mod:abspow} (simple power law
  with intrinsic absorption) superposed.  The model parameters are
  summarized in Table~\ref{tab:fits}.  {\it Right Column:} The 68, 90,
  and 99\% confidence contours in the $\Gamma$--$\nh$ plane for
  model~\ref{mod:abspow}. The 90\% upper limits to \nh\ are summarized
  in Table~\ref{tab:fits} (in the case of Q0014+8118 we are able to
  determine an intrinsic column density which is not consistent with
  zero.\label{fig:spectra}}
\end{figure*}
\begin{figure*}
\centerline{
  \hfill
  \includegraphics[angle=270,width=6.5cm]{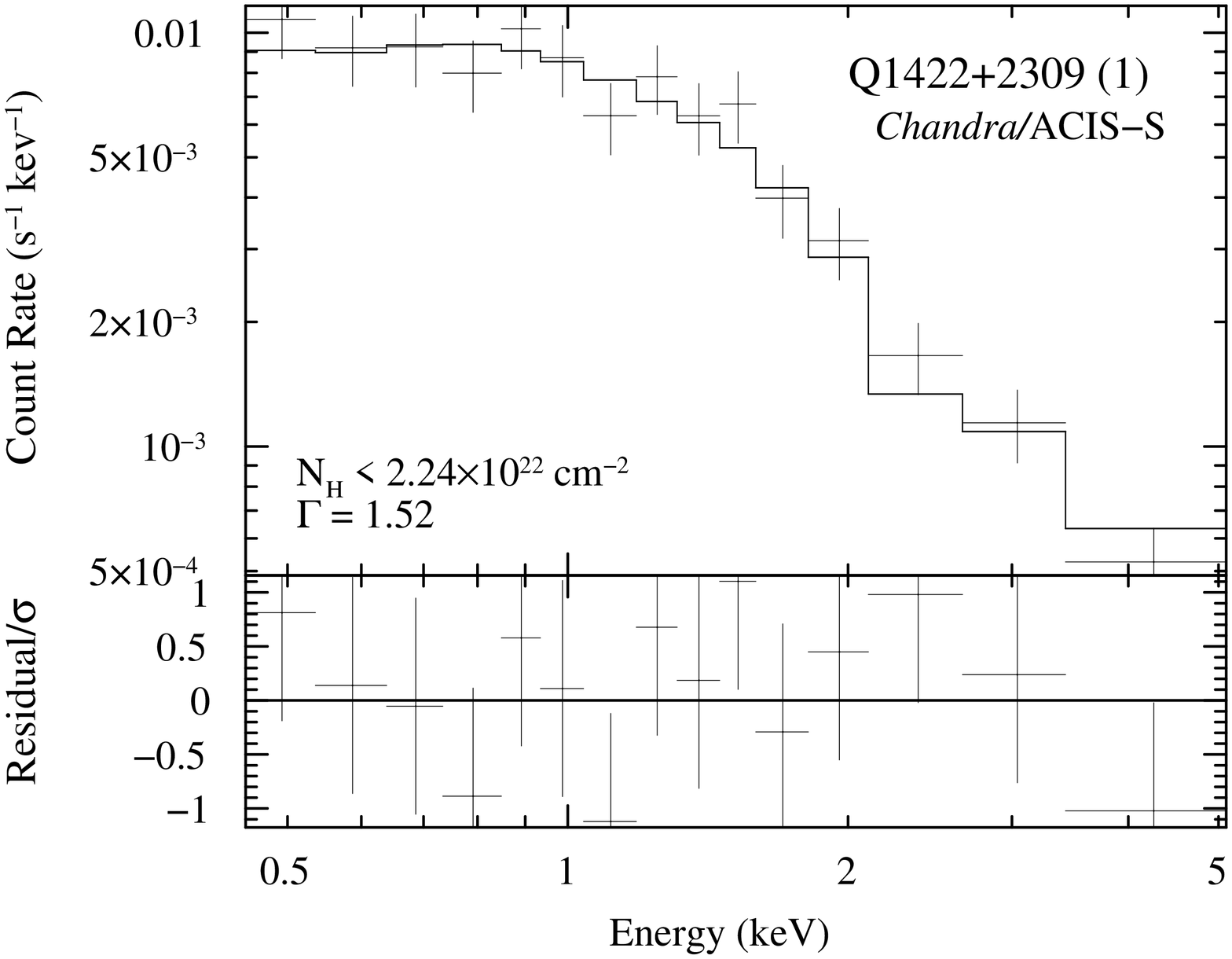}
  \hskip 0.5cm
  \includegraphics[angle=270,width=5.5cm]{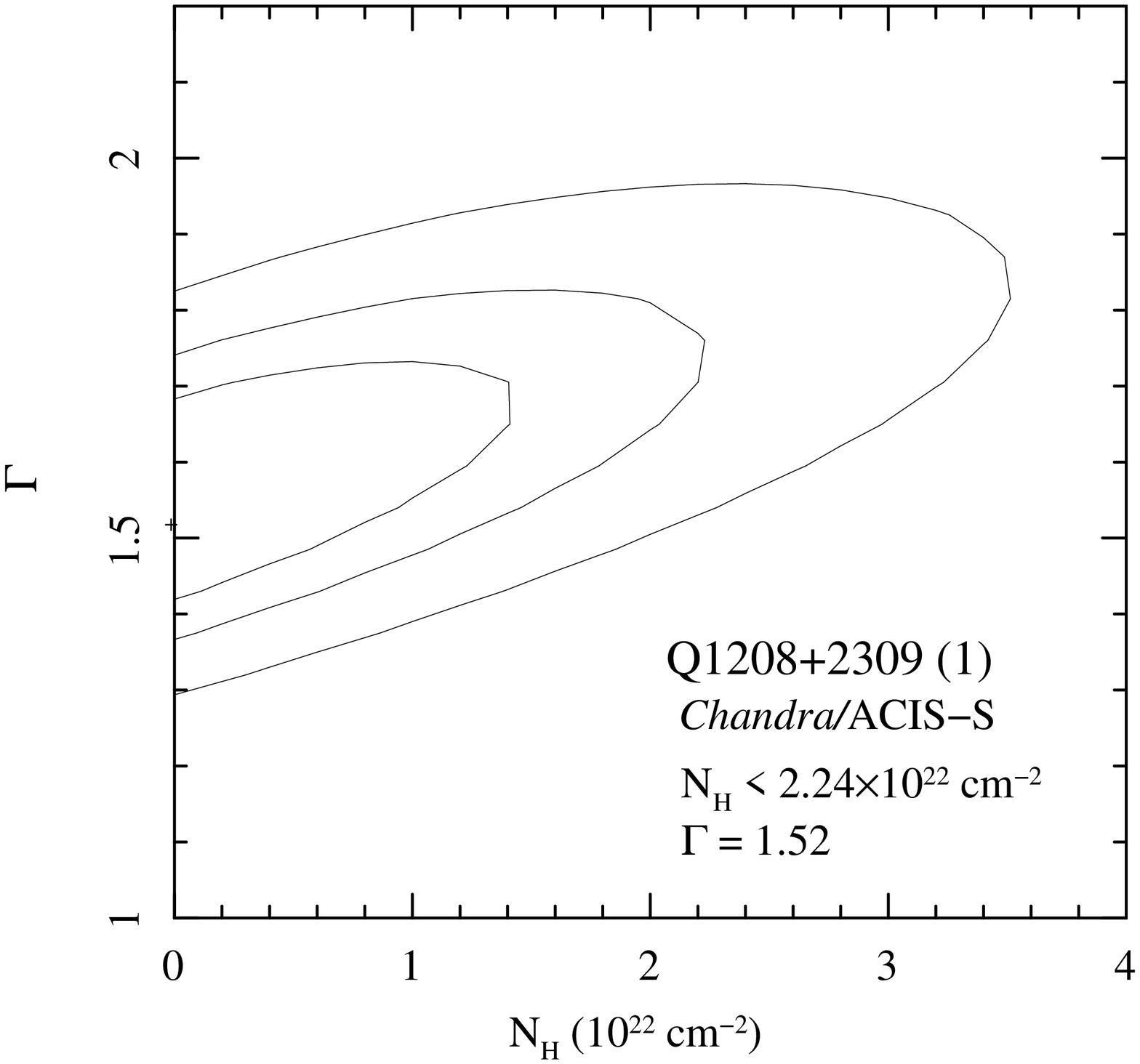}
  \hfill
}
\bigskip
\centerline{
  \hfill
  \includegraphics[angle=270,width=6.5cm]{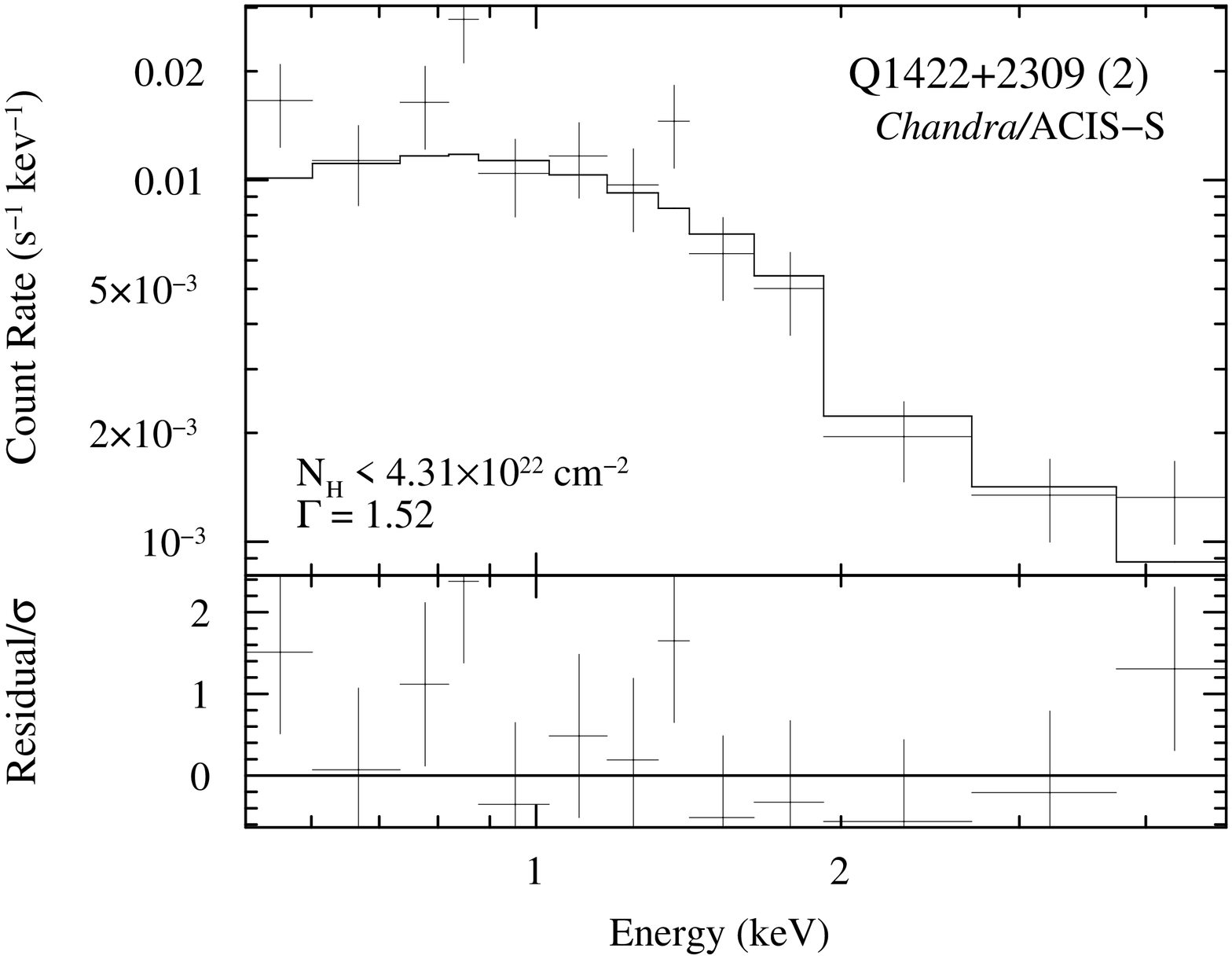}
  \hskip 0.5cm
  \includegraphics[angle=270,width=5.5cm]{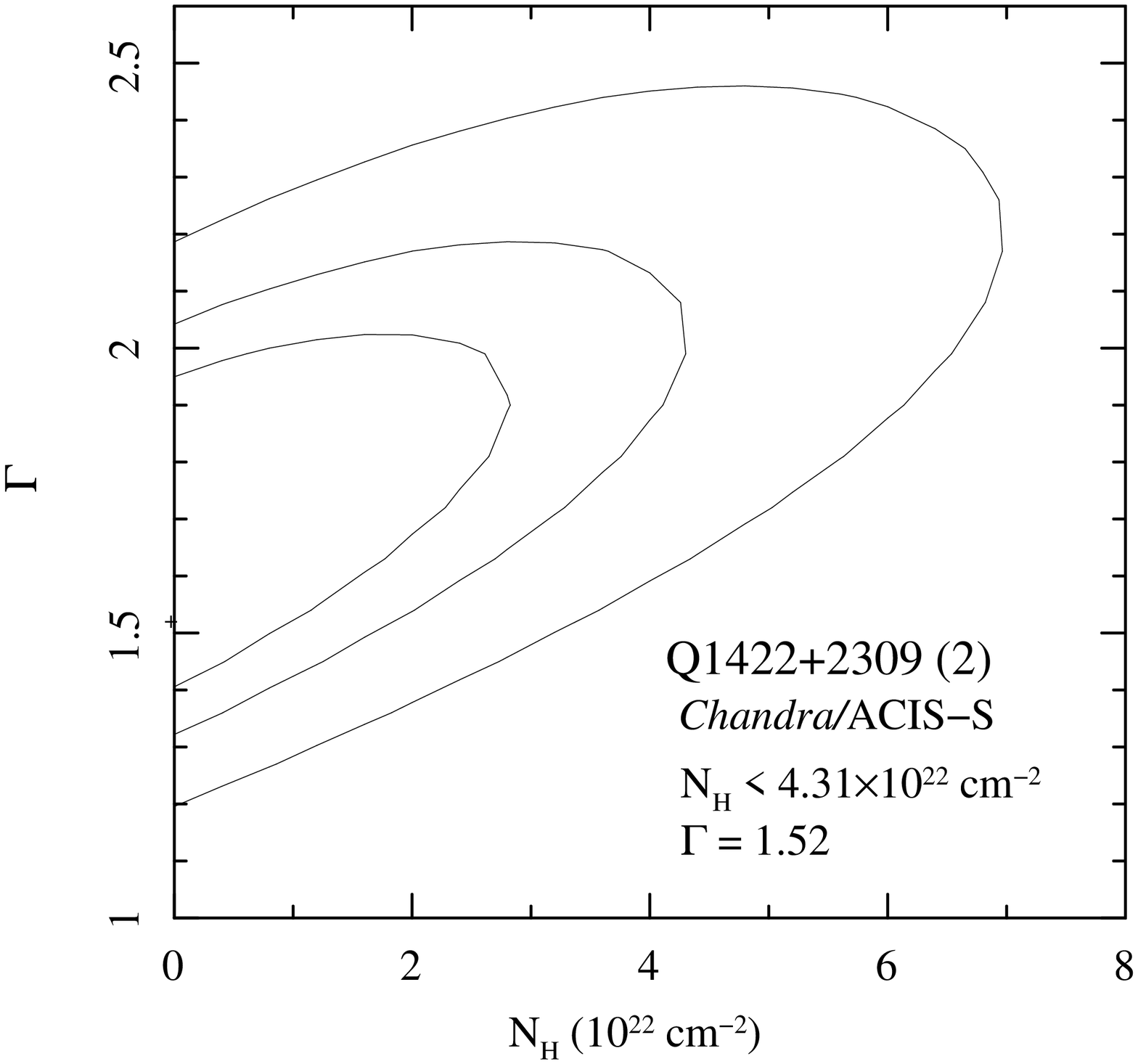}
  \hfill
}
\bigskip
\centerline{
  \hfill
  \includegraphics[angle=270,width=6.5cm]{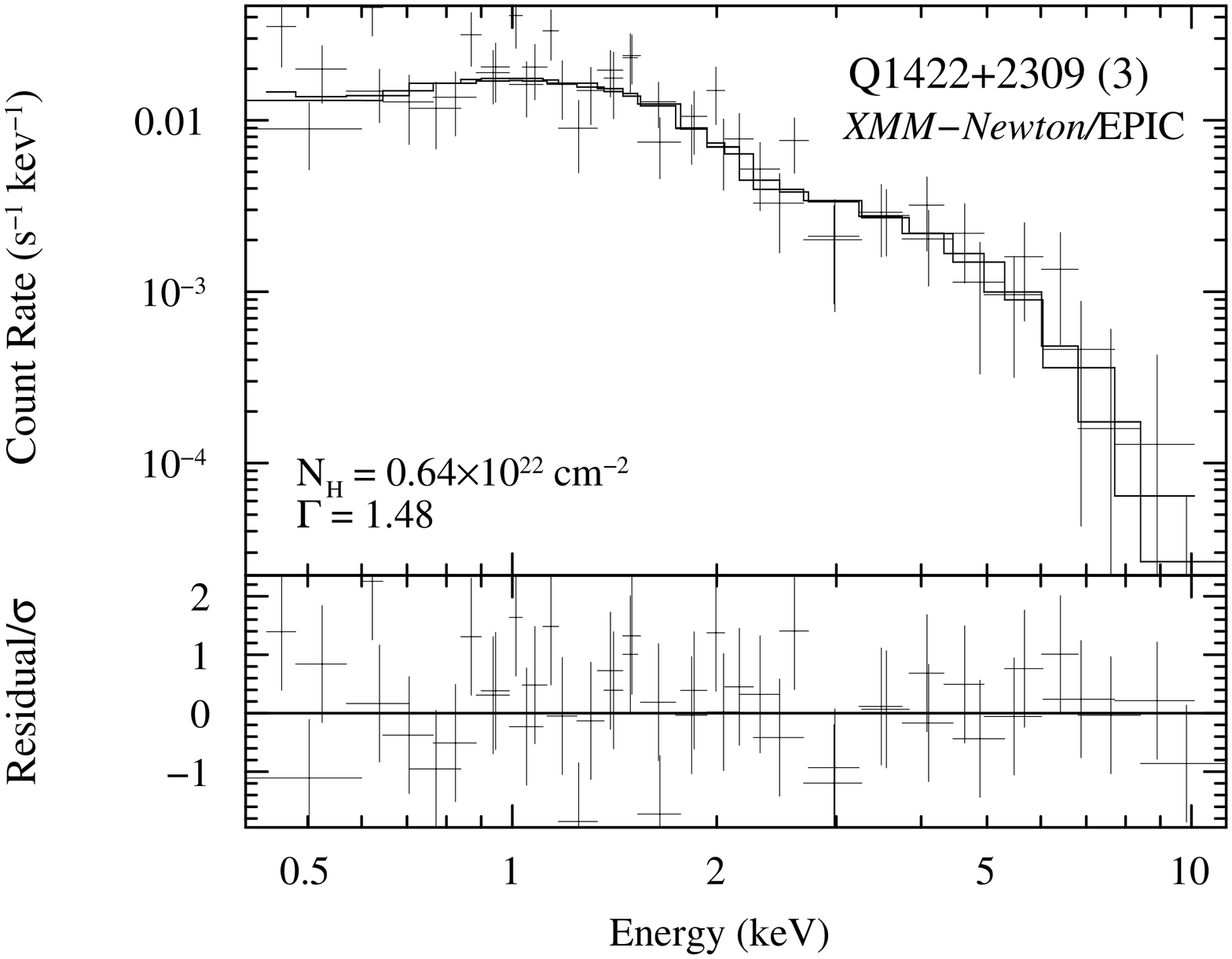}
  \hskip 0.5cm
  \includegraphics[angle=270,width=5.5cm]{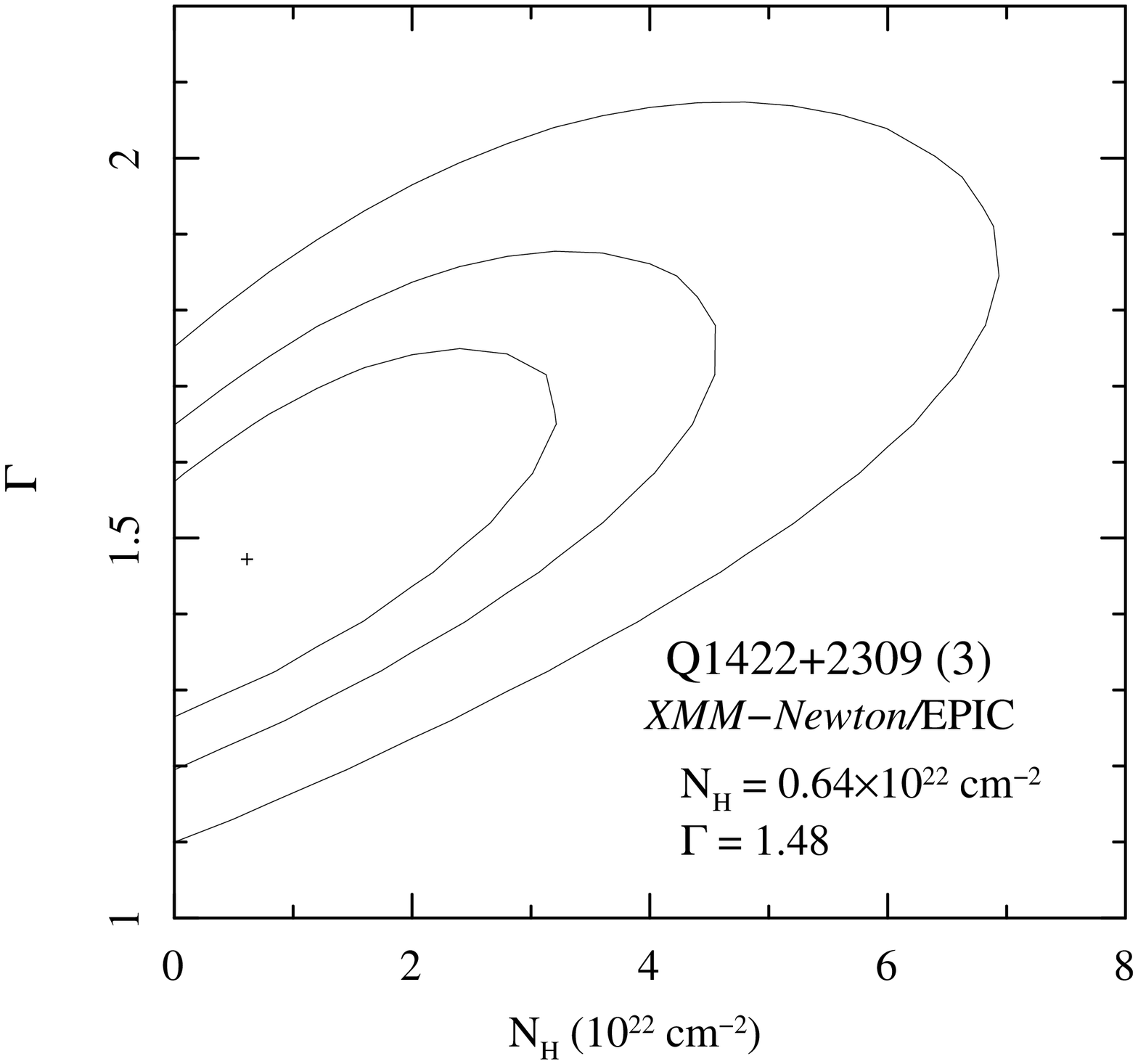}
  \hfill
}
\bigskip
\centerline{
  \hfill
  \includegraphics[angle=270,width=6.5cm]{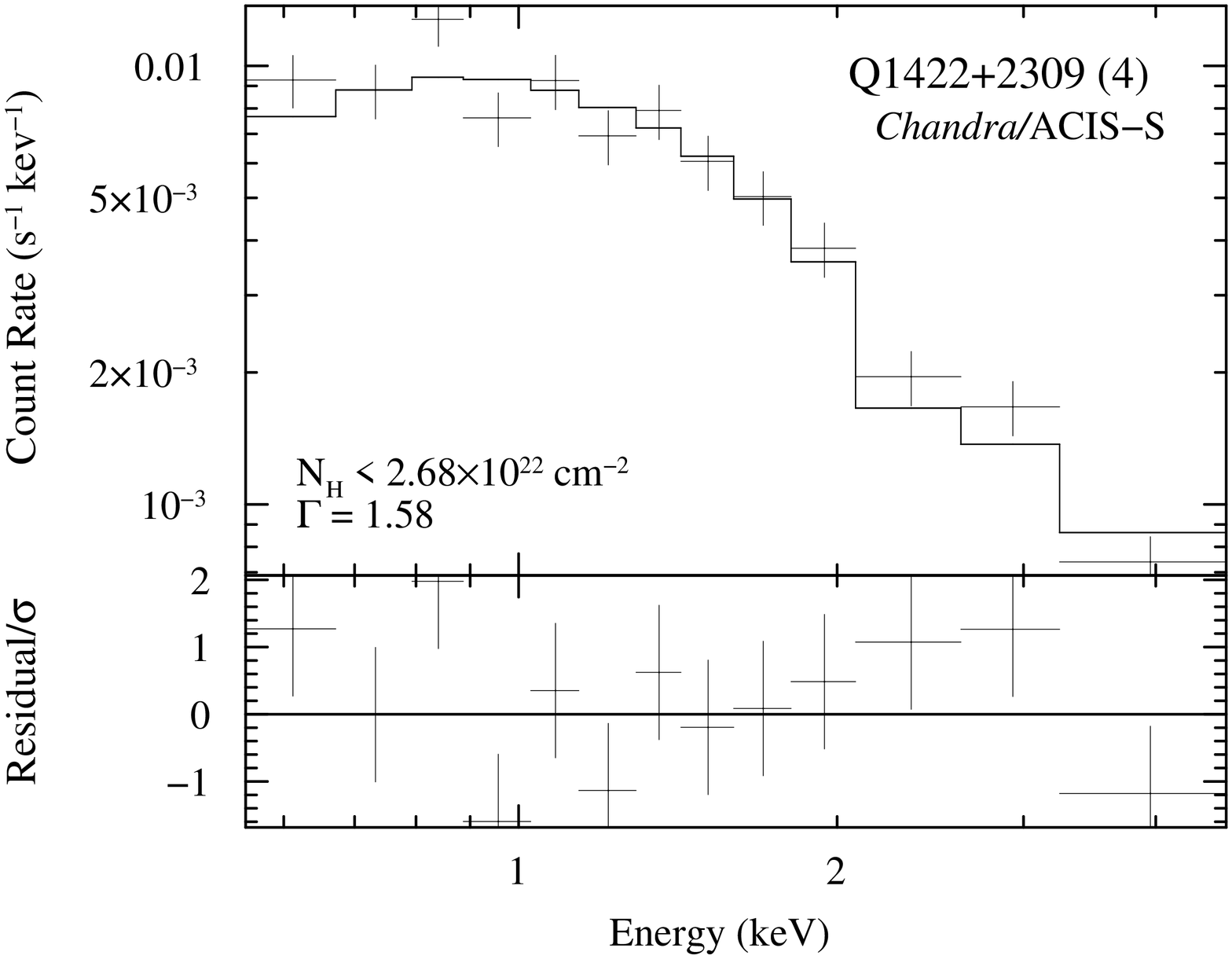}
  \hskip 0.5cm
  \includegraphics[angle=270,width=5.5cm]{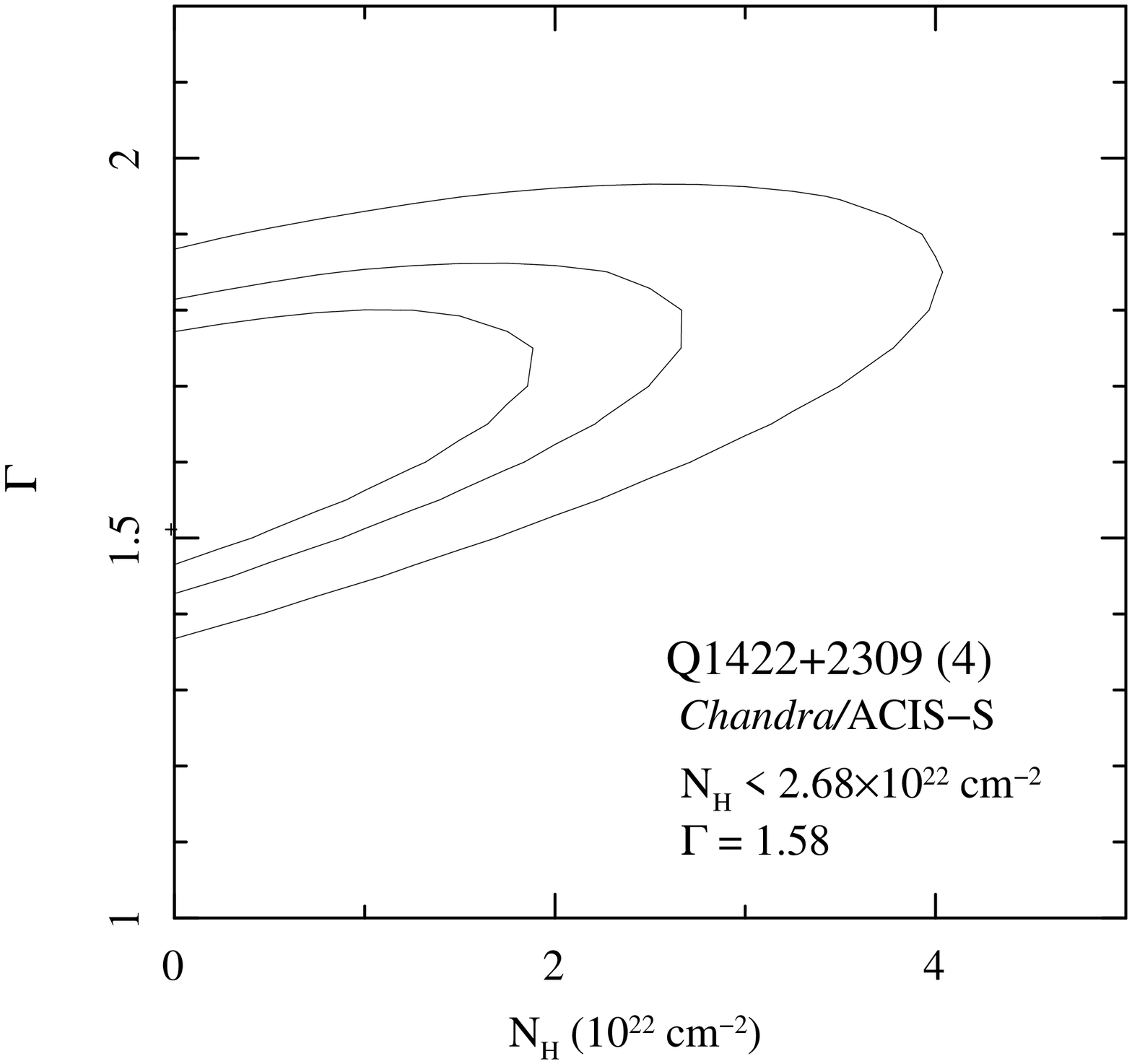}
  \hfill
}
\bigskip
\centerline{Fig. \ref{fig:spectra}.--- {\it Continued}}
\end{figure*}
\begin{figure*}
\centerline{
  \hfill
  \includegraphics[angle=270,width=6.5cm]{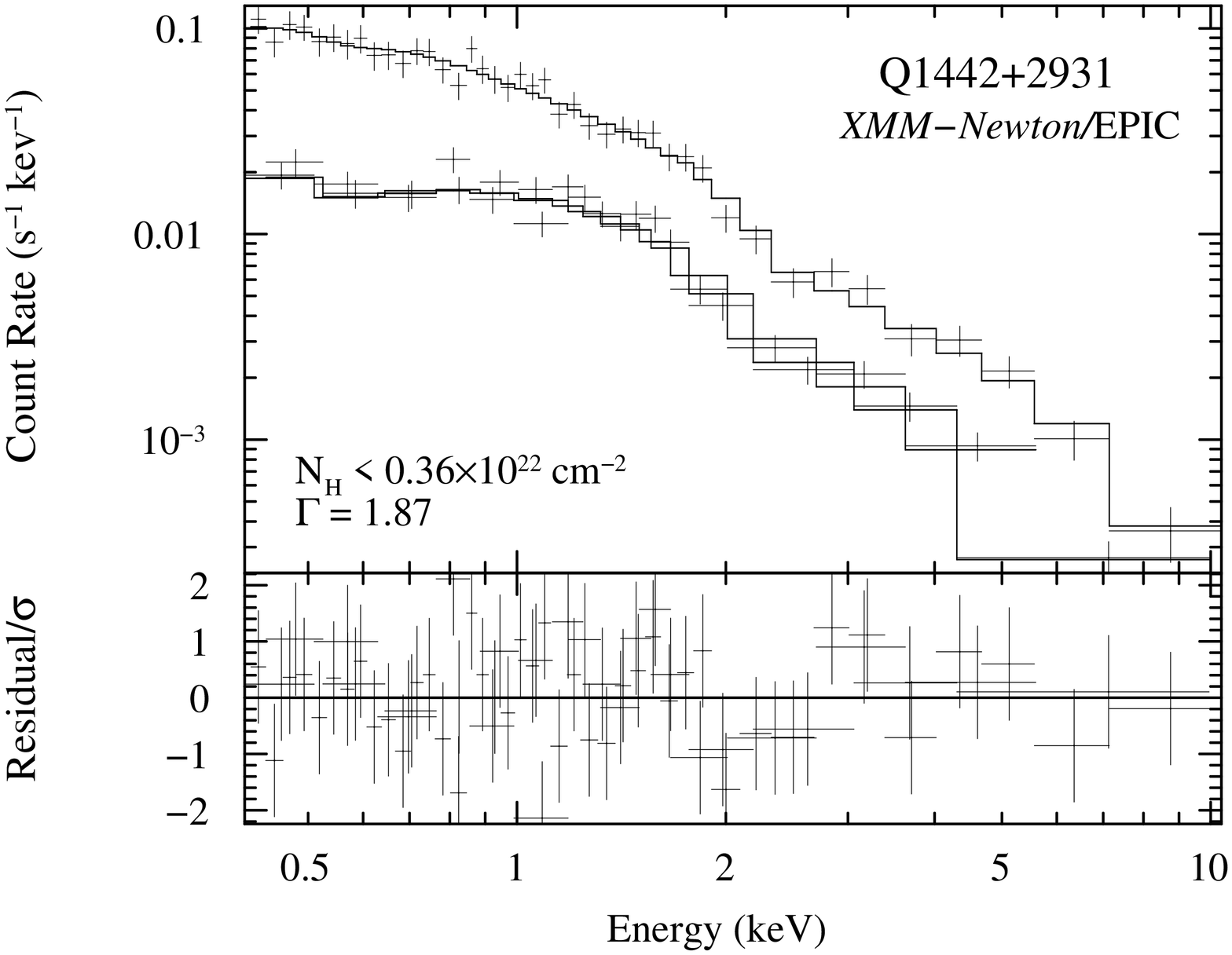}
  \hskip 0.5cm
  \includegraphics[angle=270,width=5.5cm]{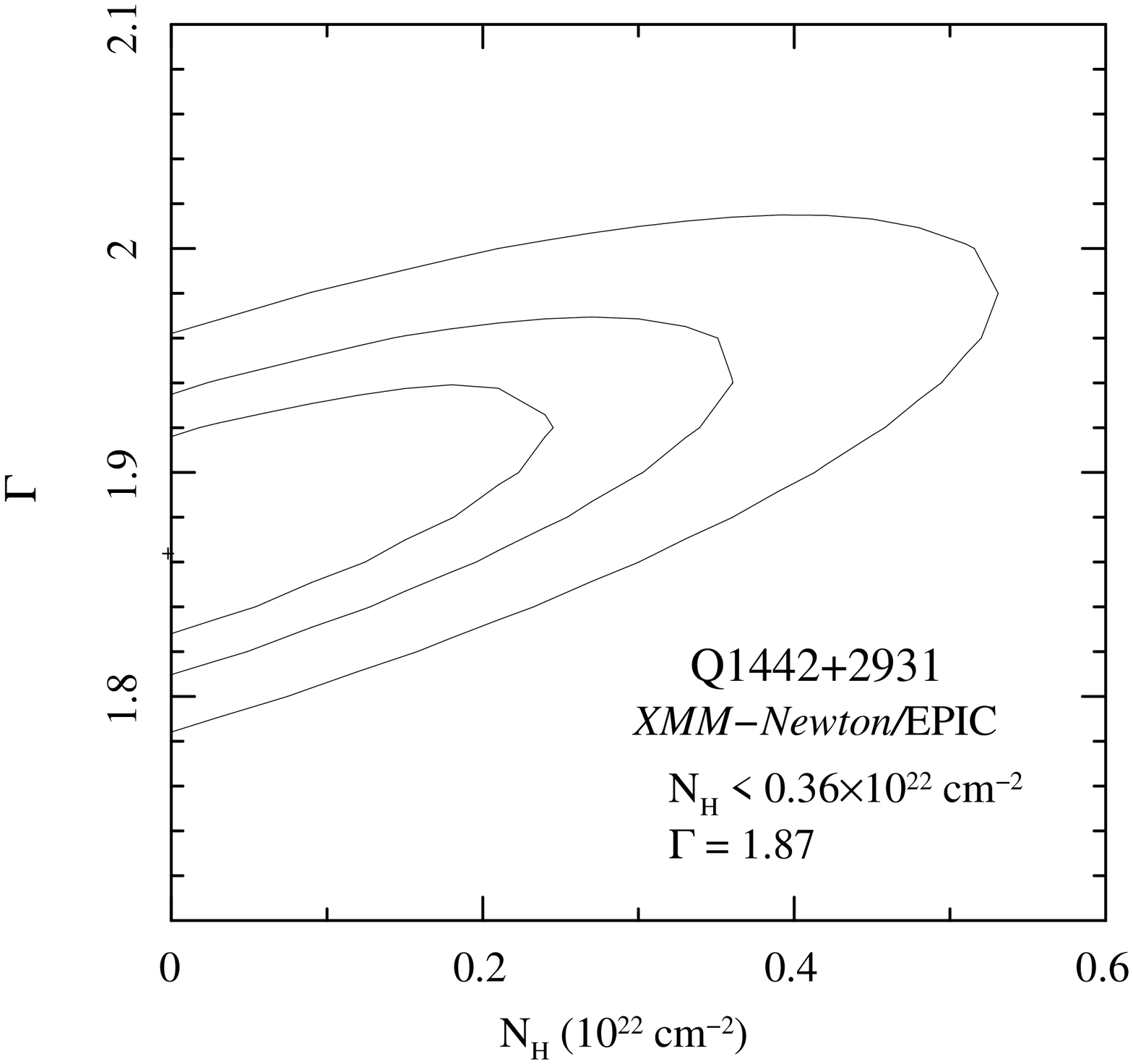}
  \hfill
}
\bigskip
\centerline{
  \hfill
  \includegraphics[angle=270,width=6.5cm]{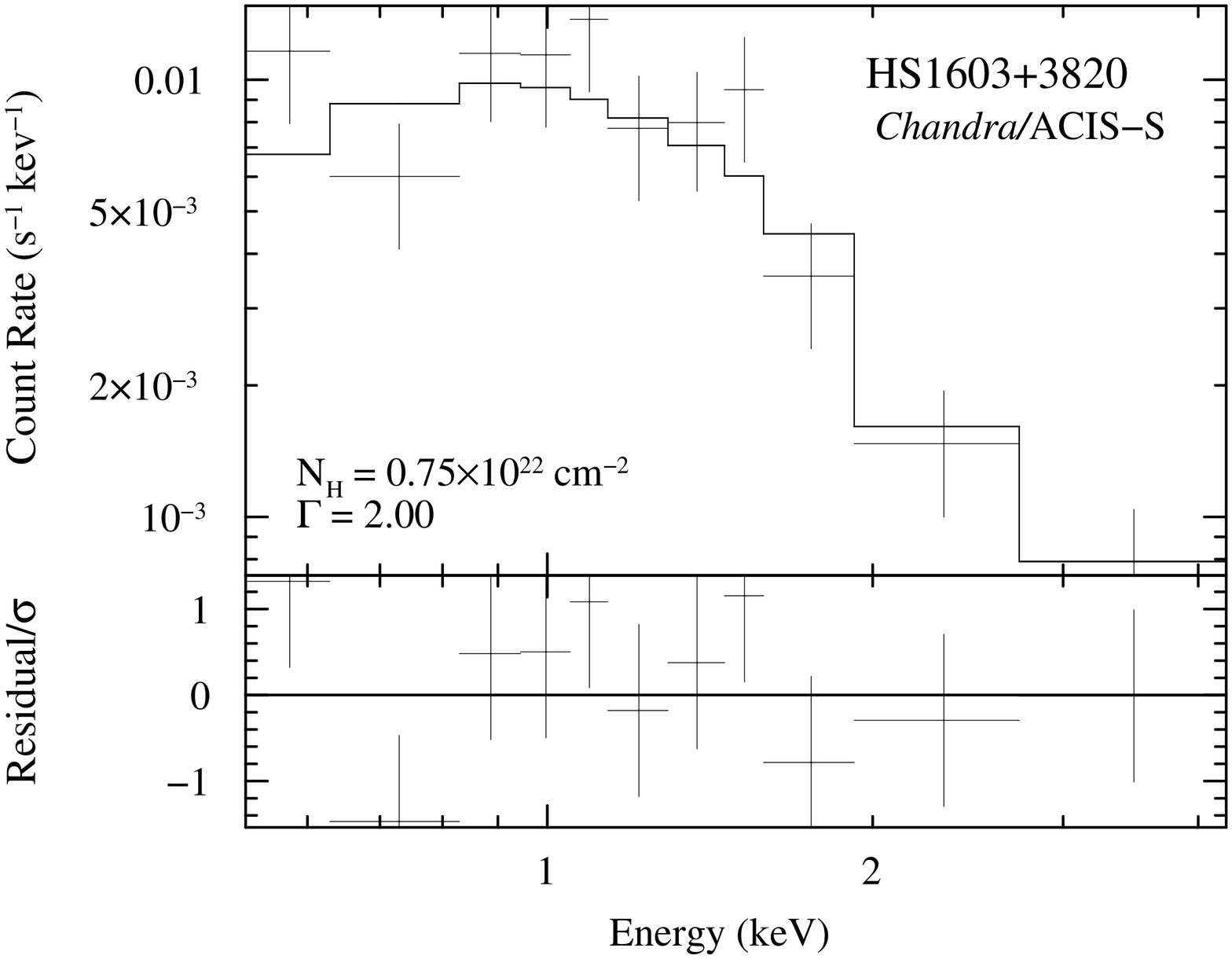}
  \hskip 0.5cm
  \includegraphics[angle=270,width=5.5cm]{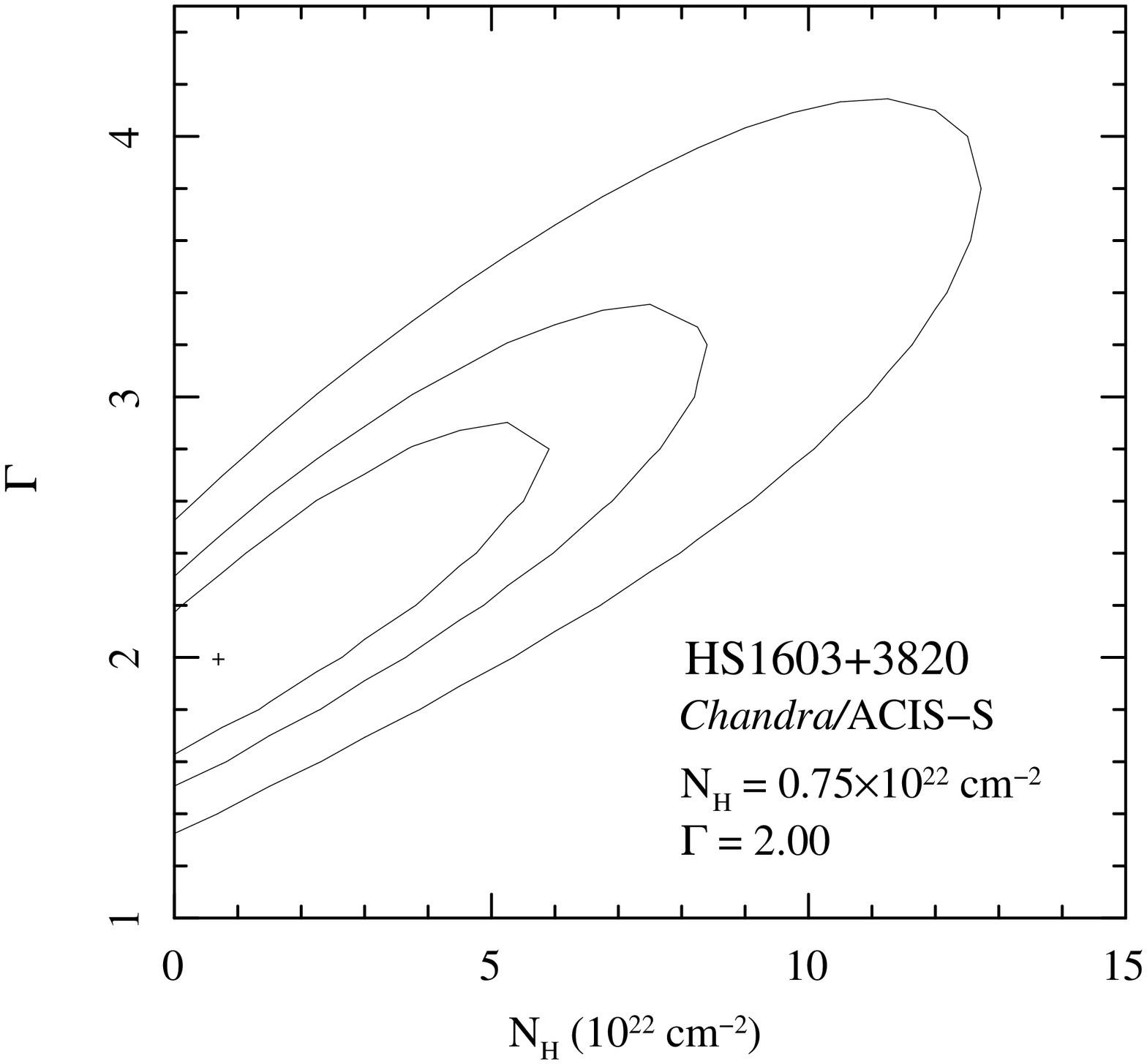}
  \hfill
}
\bigskip
\centerline{
  \hfill
  \includegraphics[angle=270,width=6.5cm]{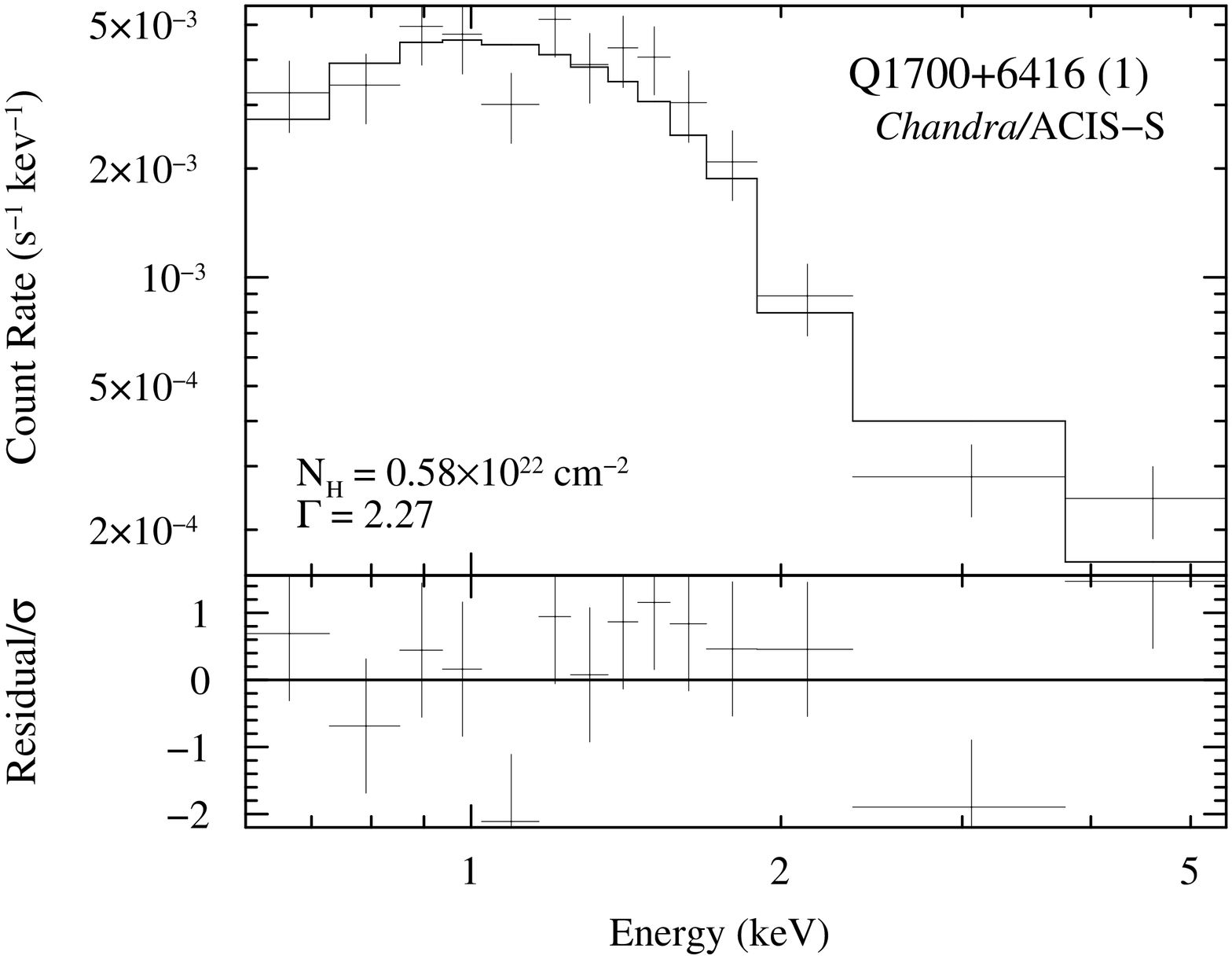}
  \hskip 0.5cm
  \includegraphics[angle=270,width=5.5cm]{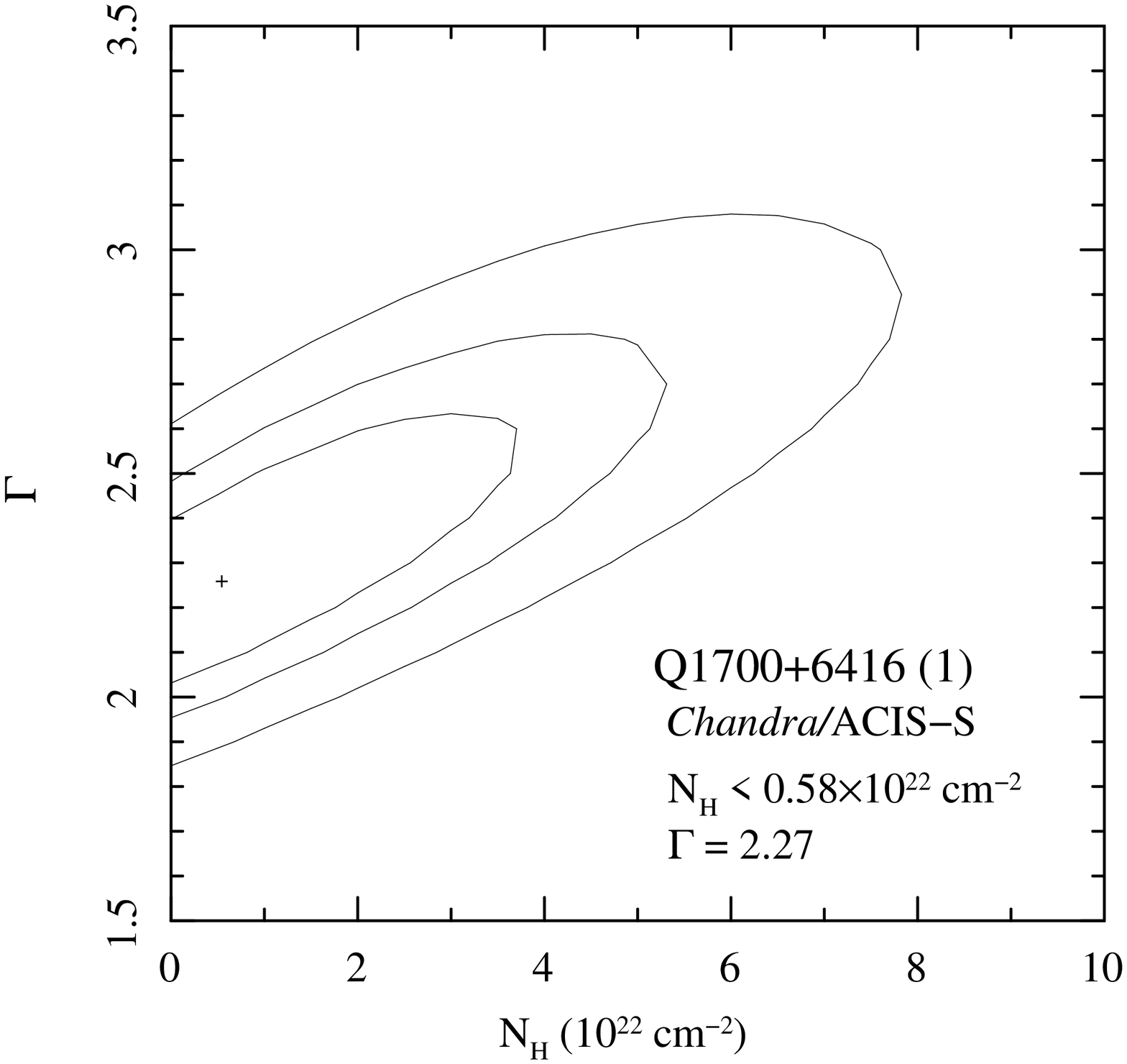}
  \hfill
}
\bigskip
\centerline{
  \hfill
  \includegraphics[angle=270,width=6.5cm]{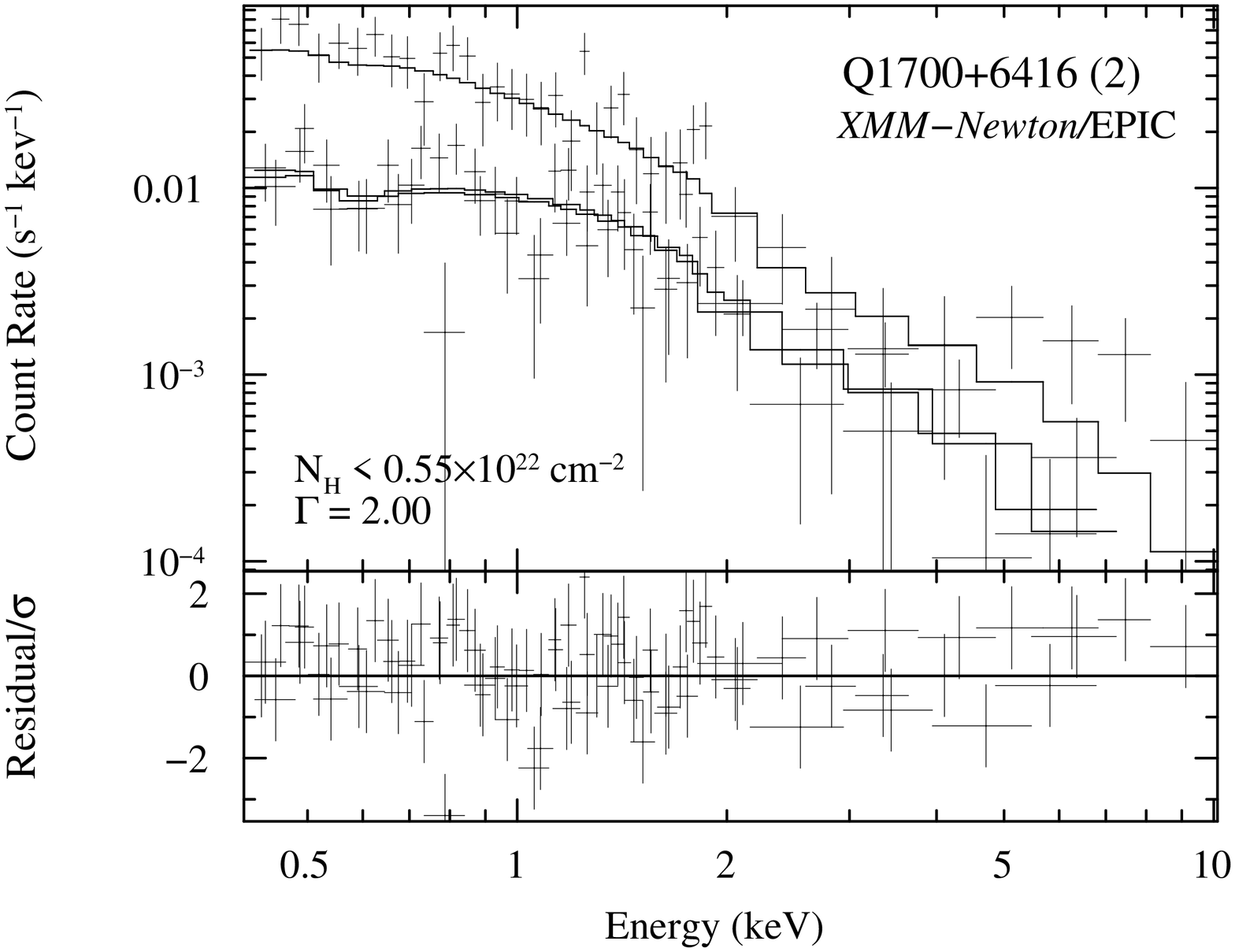}
  \hskip 0.5cm
  \includegraphics[angle=270,width=5.5cm]{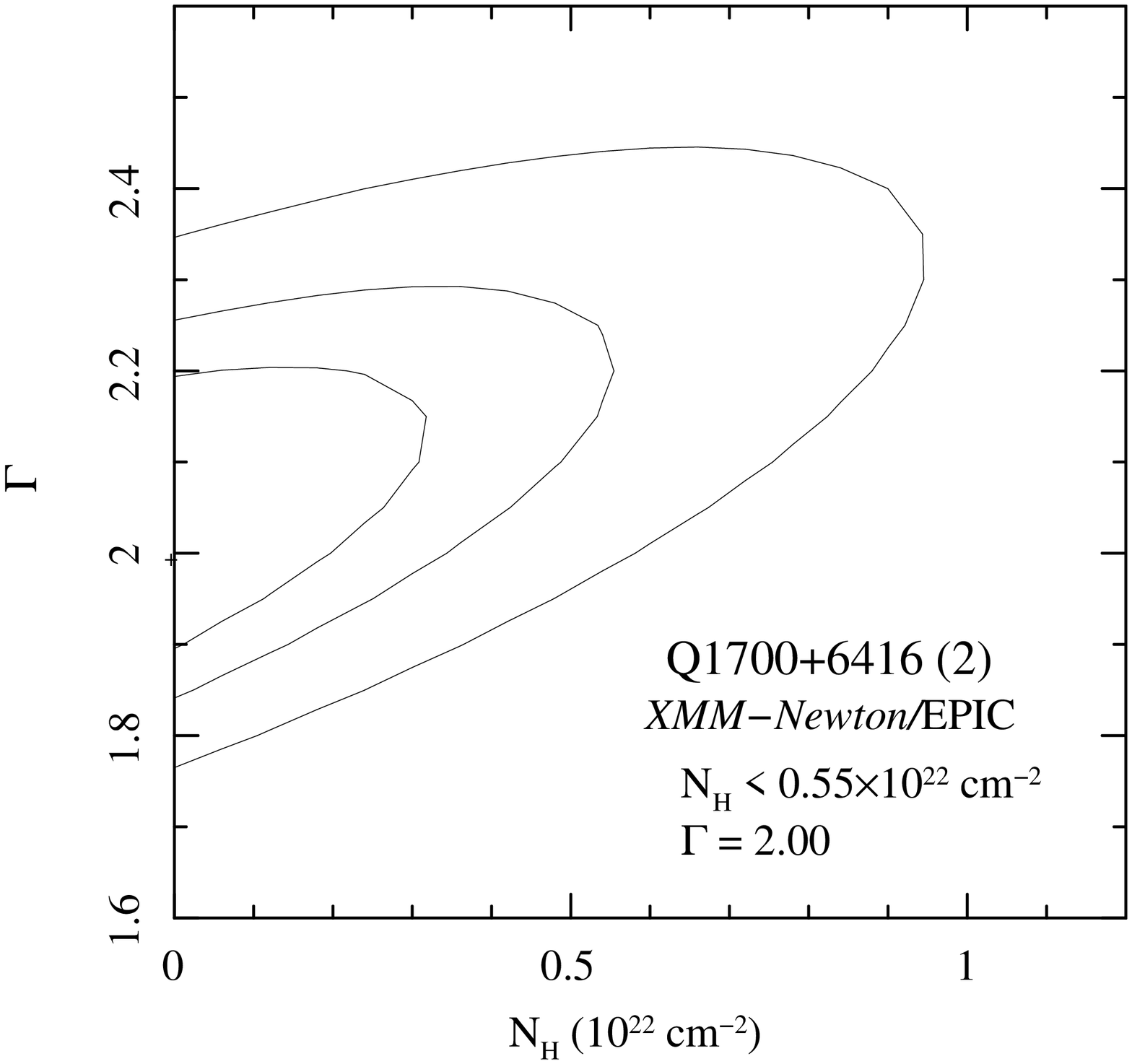}
  \hfill
}
\bigskip
\centerline{Fig. \ref{fig:spectra}.--- {\it Continued}}
\end{figure*}


We find that the simplest models from the above list provide an
adequate fit to most of the spectra, in part because of the modest
signal-to-noise ratio ($S/N$) of the spectra. In
Figure~\ref{fig:spectra}, we show the spectra with the best-fitting
power law model with intrinsic absorption (model~\ref{mod:abspow})
superposed. In the same figure, we also show the confidence contours
in the $\Gamma$--$\nh$ plane for the same model. In
Table~\ref{tab:fits} we list the best-fitting values of the photon
index for model~\ref{mod:pow} (simple power law) as well as the photon
index and intrinsic column density for model~\ref{mod:abspow} (simple
power law with intrinsic absorption).  In cases where a complex model
provides a better fit to the spectrum than the simple, unabsorbed
power law, we also list the best-fitting parameters of the complex
model. In Table~\ref{tab:iabs} we list the values (or limits) of the
column density obtained from the ionized absorber model (model
\ref{mod:absori}).  In this table, we only include those objects whose
spectra (typically from \xmm) included more than 1000 counts and
yielded ``interesting'' results. 

In 9 out of 12 spectra the simple power law without intrinsic
absorption (model~\ref{mod:pow}) provides the best fit. In another two
spectra (Q0130$-$4021 and Q1107$+$4847), the improvement in the
goodness of the fit resulting from a more complex model is only
marginal (see Table~\ref{tab:fits} and the discussion in
\S\ref{sec:notes}, below).  The remaining object is Q0014+8118, the
only case where we are able to detect significant intrinsic absorption
with a column of $\approx 1\times 10^{22}~\cmm$. This quasar happens
to be the brightest quasar in our sample by far, with more than 13,000
counts in its \xmm\ pn spectrum; we discuss this case further in
\S\ref{sec:notes} and \S\ref{sec:case}.  The best-fitting values of
the photon index are in the range 1.5--2.3, which are well within the
range of photon indices found in radio-quiet quasars at $z=2$--3
(Vignali et al. 2001).

Thus, our primary observational result consists of the upper
limits on (and one measurement of) the column density of a {\it
nearly-neutral} absorber in the rest frame of the quasar. These were
determined in a uniform and systematic fashion using the absorbed
power-law model (model~\ref{mod:abspow}) and are listed in column~5 of
Table~\ref{tab:fits}.  The limits correspond to the 90\% confidence
intervals shown in the contour plots of Figure~\ref{fig:spectra} and
are of order $10^{22}$~\cmm. The use of more complex models did not
result in more restrictive limits (most likely because of the larger
number of free parameters in these models).  Thus the limits in
column~5 of Table~\ref{tab:fits} also incorporate the uncertainty
resulting from the use of more complex models for the continuum.  We
note that these limits are considerably lower than the intrinsic
column densities detected in BAL quasars, $\nh = 10^{23-24}$~\cmm\
(e.g., Gallagher et al. 2006, and references therein).  In the case of
an {\it ionized} absorber, the data allow for higher column
densities. The corresponding limits were derived in the same fashion
as for the nearly-neutral absorber and they turn out to be
approximately one order of magnitude higher than than in the case of
the neutral absorber. They are listed in column~3 of
Table~\ref{tab:iabs}. 

\begin{figure}
\centerline{\includegraphics[angle=90,scale=0.43]{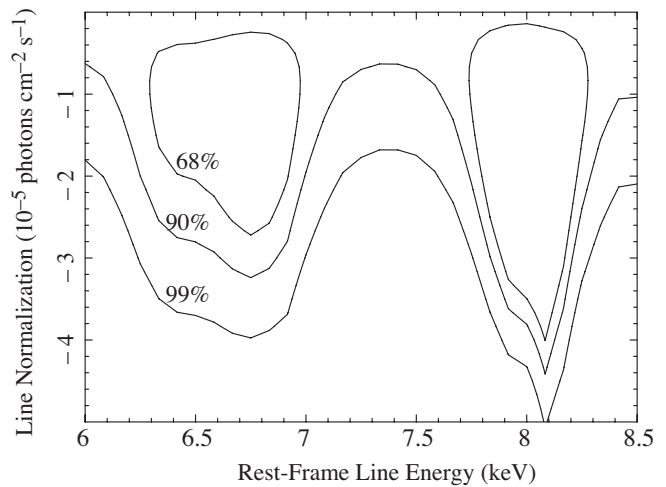}}
\caption{Limits (at 68\%, 90\%, and 99\% confidence) on the
  normalization of an unresolved Fe~H$\alpha$ absorption line in the
  spectrum of Q0014+8118. The profile of the line is assumed to be
  Gaussian and it is superposed on the power-law continuum. The line
  normalization can be translated to a rest-frame equivalent
  width using equation~(\ref{eq:Wrest}).
\label{fig:absul}}
\end{figure}

We also searched for absorption lines in the vicinity of the
Fe~K$\alpha$ transition by adding an unresolved ($\sigma=0.01$~keV)
Gaussian line with negative normalization to the power-law
continuum. This was motivated by the detection of such absorption
lines in high-redshift quasars by several authors, as we discuss
further in \S\ref{sec:prospects}. Our search yielded interesting
limits only in the case of Q0014+8118, which has by far the
highest-$S/N$ spectra of all the quasars in our sample. These limits
are shown graphically in Figure~\ref{fig:absul} where we plot the
normalization of the Gaussian component (i.e., the integral of the
Gaussian) as a function of rest-frame energy in the spectrum. Given
the redshift of this quasars and its continuum photon index, we can
relate the absolute value of the normalization of the Gaussian line,
$K$, to the rest-frame equivalent width, $W_{rest}$, analytically as
\begin{equation}
W_{rest}=26\;\left({K\over 4\times 10^{-5}{\rm\; cm^{-2}\; s^{-1}}}\right)\;
\left({E_{rest}\over 6.4\;{\rm keV}}\right)^{1.48}~{\rm eV}\; ,
\label{eq:Wrest}
\end{equation}
where $E_{rest}$ is the rest-frame energy of the absorption
line. Thus, according to Figure~\ref{fig:absul}, at a rest-frame
energy of 6.7~keV the 99\% upper limit to the absolute value of the
equivalent width is 28~eV, while at a rest-frame energy of 7.4~keV
this limit is 13~eV.

\begin{deluxetable}{rlcccrr}
\tabletypesize{\scriptsize}
\tablecaption{Flux Densities, Integrated Fluxes, and \aox\ of Sample Quasars\label{tab:flux}}
\tablewidth{0pt}
\tablehead{
\colhead{} &
\colhead{$F$(2--10 keV)\tablenotemark{a}} &
\colhead{$L$(2--10 keV)\tablenotemark{b}} &
\colhead{\fx\tablenotemark{c}} &
\colhead{} &
\colhead{} \\
\colhead{Quasar} & 
\colhead{(${\rm erg~s^{-1}~cm^{-2}}$)} &
\colhead{(erg~s$^{-1}$)} &
\colhead{(nJy)} &
\colhead{\aox\tablenotemark{d}} &
\colhead{\daox\tablenotemark{e}} \\
\colhead{(1)} & 
\colhead{(2)} & 
\colhead{(3)} &
\colhead{(4)} &
\colhead{(5)} &
\colhead{(6)} 
}
\startdata
Q0014$+$8118  & $3.6\times 10^{-12}$        & $1.5\times 10^{47}$ & 126   & $-1.44$ & $ 0.42$ \\
Q0130$-$4021  & $8.3\times 10^{-14}$        & $5.8\times 10^{45}$ & 11.3  & $-1.63$ & $ 0.14$ \\
Q1107$+$4847  & $3.2\times 10^{-14}$        & $2.0\times 10^{45}$ & 3.65  & $-1.88$ & $-0.09$ \\
Q1208$+$1011  & $3.2\times 10^{-14\;\rm f}$ & $1.7\times 10^{45}$ & 2.36  & $-1.61$ & $ 0.11$ \\
Q1422$+$2309  & $7.1\times 10^{-14\;\rm g}$ & $1.1\times 10^{45}$ & 0.93  & $-1.81$ & \dots   \\
              & $9.4\times 10^{-14\;\rm g}$ & $1.7\times 10^{45}$ & 1.61  & $-1.72$ & \dots   \\
              & $8.7\times 10^{-13\;\rm f}$ & $2.4\times 10^{45}$ & 1.86  & $-1.69$ & \dots   \\
              & $8.3\times 10^{-14\;\rm g}$ & $1.4\times 10^{45}$ & 1.29  & $-1.75$ & $-0.03$ \\
Q1442$+$2931  & $1.6\times 10^{-13}$        & $7.0\times 10^{45}$ & 16.5  & $-1.70$ & $ 0.10$ \\
HS1603$+$3820 & $5.8\times 10^{-14}$        & $2.3\times 10^{45}$ & 6.1 & $-1.96$ & $-0.15$ \\
Q1700$+$6416  & $2.5\times 10^{-14}$        & $1.7\times 10^{45}$ & 5.06  & $-1.92$ & \dots   \\
              & $8.6\times 10^{-14}$        & $4.8\times 10^{45}$ & 12.5  & $-1.77$ & $-0.04$ \\
\enddata
\tablenotetext{a}{The 2--10~keV flux in the observed frame,
                    corrected for absorption in the ISM of the Milky
                    Way (and for intrinsic absorption in the case of
                    Q0014$+$8118).}
\tablenotetext{b}{The rest-frame 2--10~keV luminosity, corrected for
                    absorption and scaled down to account for the
                    lensing magnification in the case of the two
                    lensed quasars, Q1208$+$1011 and Q1422$+$2309.}
\tablenotetext{c}{The rest-frame flux density per unit frequency at
                    an energy of 2~keV (corrected for Milky Way
                    absorption but not intrinsic absorption; see
                    \S\ref{sec:xuv}). The uncertainties are $\ls
                    10$\%.}
\tablenotetext{d}{The optical-to-X-ray spectral index computed from
                    equation(\ref{eq:defaox}) {\it without}
                    corrections for intrinsic absorption. The
                    uncertainties are discussed in \S\ref{sec:xuv} of
                    the text.}
\tablenotetext{e}{The difference between the observed value of \aox\
                    and the value expected based on the UV luminosity
                    of the quasar (see details in \S\ref{sec:xuv} of
                    the text). For quasars with multiple X-ray
                    observations, we only report the value of \daox\
                    for the observation with the highest $S/N$.}
\tablenotetext{f}{Total observed flux of all lensed images combined,
                    not corrected for magnification (see notes in
                    \S\ref{sec:notes}).}
\tablenotetext{g}{Observed flux of image C only, not corrected for
                    magnification (see notes in \S\ref{sec:notes}).}
\end{deluxetable}

\subsection{Notes on Individual Quasars}\label{sec:notes}

\begin{description}

\item[{\it Q0014+8118. ---}] Page et al. (2004) presented the \xmm\
   spectrum of this quasar in their analysis of the X-ray Baldwin
   effect. We reduced the same \xmm\ data taken on 2001 August 23 for
   14.3~ks (pn) and 16.9~ks (MOS) (Obs-ID: 0112620201, PI:
   Turner). The spectrum was binned with a minimum of 50 counts per
   bin. The rest-frame 2--10~keV luminosity that we obtain is in good
   agreement with that of Page et al. (2004); they do not report any
   of the parameters of the best-fitting model. This is the only
   object in which we detect an intrinsic absorber unambiguously, with
   a column of $\approx 1\times 10^{22}$~\cmm.  The fully and
   partially covered, neutral absorber models, and the ionized
   absorber model (models~\ref{mod:abspow}, \ref{mod:pcovpow}, and
   \ref{mod:absori}) yield a significantly better fit than the simple
   power-law model (model~\ref{mod:pow}), as shown by the F-test (see
   chance probabilities in Table~\ref{tab:fits}).  The ionized
   absorber model of Done et al. (1992), as implemented in XSPEC, is
   not self-consistent, therefore its results must be regarded with
   caution.  In particular, the temperature and ionization parameter
   are treated as independent parameters instead of being calculated
   self-consistently from a full photoionization model. Therefore, we
   verified the results using a rigorous photoionization model,
   computed by the code XSTAR\footnote{\tt
   http://heasarc.gsfc.nasa.gov/docs/software/xstar/xstar.html}
   (Kallman et al. 2004). The latter model yielded a column density
   for the ionized absorber of $N_{\rm H}=(2.5\pm 2.0)\times
   10^{22}\;{\rm cm^{-2}}$ and an ionization parameter of $\log
   \xi_{\rm XSTAR} = 1.7\pm 0.7$ (where $\xi$ is in units of
   erg~cm~s$^{-1}$). In comparison, the fit with model
   \ref{mod:absori} (\verb+absori+) yields $N_{\rm
   H}=(1.5^{+3.4}_{-0.6})\times 10^{22}\;{\rm cm^{-2}}$ and an
   ionization parameter of $\log\xi_{\tt absori} < 2.2$, corresponding
   to $\log \xi_{\rm XSTAR} < 1.9\;$\footnote{The ionization parameters
   in the XSTAR and {\tt absori} models are defined in a slightly
   different way. The former involves the continuum luminosity from
   1~Ry to 1000~Ry, while the latter involves the continuum luminosity
   from 5~eV to 300~keV. They can be related if one assumes a specific
   shape for the spectral energy distribution. Here we make the
   assumption that the spectral energy distribution is that of Mathews
   \& Ferland (1987) but we change the X-ray photon index to 1.49 so
   that it agrees with what we measure for Q0014+8118. Under this
   assumption we find that $\log\xi_{\tt absori} - \log\xi_{\rm XSTAR}
   = 0.262$.}, consistent with the XSTAR results.

   Models~\ref{mod:abspow}, \ref{mod:pcovpow}, and \ref{mod:absori},
   as well as the XSTAR-based model, appear to provide equally good
   fits to the spectrum, indicating that excess absorption is present
   in the spectrum of Q0014+8118 but the properties of the ionization
   state and coverage fraction of the absorber are not well
   constrained.  Thus, we adopt model~\ref{mod:abspow} as the working
   model for this object in the discussion below because it is the
   simplest one. However, we return to this case in \S\ref{sec:xuv}
   and \S\ref{sec:discussion} to discuss the implications of the
   measured column density and to constrain the location and
   properties of the absorber with the help of additional evidence.

\item[{\it Q0130-4021. ---}] Page et al. (2003) searched for AGNs in
   the \xmm\ field of this quasar, and detected 9 objects including
   Q0130$-$4021 itself. We reduced the same \xmm\ data taken on 2001
   June 04 for 23.0~ks (pn) and 25.2~ks (MOS) (Obs-ID: 0112630201, PI:
   Turner). The pn and MOS spectra were binned with a minimum of 50
   and 30 counts per bin, respectively. Model~\ref{mod:compt} (power
   law plus Compton reflection) provides a marginally better fit than
   model~\ref{mod:pow} (simple power-law model) to the \xmm\ spectrum
   of this object, $P_F(1)=0.0011$ (see
   Table~\ref{tab:fits}). However, the fit with model~\ref{mod:compt}
   appears to be as good as the fit with model~\ref{mod:abspow}
   (absorber power-law).

\item[{\it Q1107+4847. ---}] \xmm\ observed this target twice on 2002
   April 25 and June 1.  The analysis of the second data set was
   presented by Brocksopp et al. (2004). They fitted the spectra with
   a power law model with photon index of $\Gamma$ = 1.96 and obtained
   an upper limit on the intrinsic absorption of $\nh <
   1.5\times10^{22}$~\cmm\ with 99\%\ confidence.  We examined both
   \xmm\ data sets (Obs-ID: 0059750401 and 0104861001, PI: Mann and
   Mason), but reduced only the second one because the first
   observation was performed during a period of high background
   (background-to-signal ratio is $>$20).  The pn and MOS spectra were
   binned with a minimum of 50 and 15 counts per bin,
   respectively. The results of our fits are consistent with the
   results of Brocksopp et al. (2004). Model~\ref{mod:abscompt} (power
   law plus Compton reflection and intrinsic absorption) provides an
   equally good fit to the \xmm\ spectrum of this object as
   model~\ref{mod:pow} (simple power-law model) and
   model~\ref{mod:abspow} (absorber power-law).

\item[{\it Q1208+1011. ---}] This gravitationally lensed quasar was
   originally found by Hazard, McMahon, \& Sargent (1986). It
   comprises two split images separated by 0{\farcs}47 with a
   brightness ratio of about 4, based on the observations with
   ground-based telescopes (Magain et al. 1992) and the Hubble Space
   Telescope (HST) (Bahcall et al. 1992b). Giallongo et al. (1999)
   derived a total magnification factor of $\mu_t=22$, which was later
   revised to $\mu_t = 3.1$ by Barvainis \& Ivison (2002). We
   extracted the spectrum of this quasar by summing the counts from
   all images and then scaled down the luminosity according to the
   latter magnification factor listed above.

   Dai et al. (2004) analyzed the \chandra\ data of this quasar and
   fitted the spectrum with a power law and a Gaussian emission line
   without intrinsic absorption. They found a best-fitting photon
   index of $\Gamma = 2.3\pm 0.2$ and derived value of $\aox =
   -1.62$. We repeated the reduction of the same \chandra\ data
   (Obs-ID: 3570 PI: Garmire). The spectrum was binned with a minimum
   of 10 counts per bin.  The spectral parameters that we derive are
   consistent to those obtained by Dai et al. (2004).

\item[{\it Q1422+2309. ---}] This is a gravitationally-lensed,
   radio-loud quasar, consisting of 4 images. Kormann, Schneider, \&
   Bartelmann (1994) explored lensing models based on elliptical mass
   distribution and derived a total magnification factor of $\mu_t =
   15.38$.  Three \chandra\ (Obs-ID: 367, 1631, and 4939 PI: Garmire)
   and \xmm\ observations (Obs-ID: 0143652301 PI: Georgantopoulos)
   were carried out over a span of four years.  Among the four lensed
   images resolved by \chandra, the brightest two images are probably
   affected by pileup at least in their central few pixels (Grant et
   al. 2004). Since the faintest image is not bright enough to be
   detected in the \chandra\ data, we extracted a spectrum only from
   image~C, the third brightest image.  The magnification factor for
   this particular image, according to Kormann et al. is $\mu_C =
   3.43$, thus we scaled down the luminosity accordingly.  The spectra
   from the three \chandra\ observations were binned with a minimum of
   25, 15, and 50 counts per bin, respectively. In our analysis of the
   \xmm\ observation, we used only the MOS data because in the pn the
   image fell on a chip gap. We summed all the counts from the MOS
   images since the four lensed images were not resolved. We binned the
   extracted spectrum with a minimum of 10 counts per bin and we
   scaled down the luminosity by a factor of $\mu_{tot} = 15.38$.

\item[{\it Q1442+2931. ---}] An analysis of the \xmm\ spectrum of
   this quasar was presented by Ferrero \& Brinkmann (2003). They
   fitted with a power-law and obtained a photon index of 1.87. They
   found no evidence of an Fe~H$\alpha$ line.  We re-analyzed the same
   \xmm\ data (Obs-ID: 0103060201, PI: Aschenbach). The spectra were
   binned with a minimum of 50 counts per bin. Our derived values of
   the photon index and of \aox\ are consistent with those of Ferrero
   \& Brinkmann (2003).

\item[{\it HS1603+3820. ---}] The short \chandra\ observation of this
   quasar yielded only 137 counts. The spectrum was binned with a
   minimum of 10 counts per bin. An independent analysis of the same
   X-ray data by Dobrzycki et al. (2007) yields results consistent
   with ours.

\item[{\it Q1700+6416. ---}] There are two clusters of galaxies
   discovered near the quasar, RX~J1701.3$+$6416 at $z$ = 0.45 and
   Abell 2246 at $z$ = 0.22. Both the \chandra\ and the \xmm\ observed
   this quasar field for a purpose of studying these clusters (Lumb et
   al. 2004; Georgantopoulos \& Georgakakis 2005) and serendipitously
   detected a number of low-redshift galaxies in the field
   (Hornschemeier et al. 2005). We reduced the \chandra\ data (Obs-ID:
   547, PI: Vanspeybroeck) and the \xmm\ data (Obs-ID: 0107860301, PI:
   Jansen) and we present the spectrum of Q1700+6416 for the first
   time. Although the \xmm\ exposure was terminated prematurely and
   most of it was dominated by high background from the particle belt
   (Lumb et al. 2004), we were still able to extract spectra of
   acceptable quality. The \chandra\ spectrum was binned with a
   minimum of 20 counts per bin.  The \xmm\ pn and MOS spectra were
   binned to a minimum of 50 and 15 counts per bin, respectively, at
   $E < 2$~keV and to a minimum of 150 and 45, respectively at $E >
   2$~keV.

\end{description}

\section{Relations Between UV and X-ray Properties}\label{sec:xuv}

The relation between the UV and X-ray luminosity of quasars, is often
quantified by means of the optical-to-X-ray spectral index, \aox. This
is the index of a power law connecting the points in the rest-frame
spectral energy distribution at 2500~\AA\ and 2~keV. Here we adopt the
convention that $f_{\nu}\propto\nu^{\aox}$ and we obtain \aox\ from
\begin{equation}
\alpha_{\rm ox} \equiv {\log \fx - \log \fuv \over
\log \nu_{\rm 2\; keV} - \log \nu\raise 0.3 em \hbox{$_{\scriptstyle 2500\;\sang}$}}
= 0.384\; \log \left({\fx \over \fuv}\right)\;
\label{eq:defaox}
\end{equation}
where \fx, \fuv, $\nu_{\rm 2\; keV}$, and $\nu\raise 0.3 em
\hbox{$_{\scriptstyle 2500\;\sang}$}$ are the flux densities per unit
frequency and frequencies at 2~keV and 2500~\AA, respectively. The
above definition was originally introduced by Tananbaum et al. (1979)
but several authors since then have used the flux density at 3000~\AA\
instead of 2500~\AA\ (e.g., Brandt et al. 2000; Gallagher et
al. 2001). In this paper, we adopt the frequencies used in the
original definition. The values of \aox\ obtained from the two
different conventions are related by $\aox({\rm 2500\;\AA}) = 1.03\;
\aox({\rm 3000\;\AA})- 0.03\;\alpha_{\rm UV}$, where $\alpha_{\rm UV}$
is the power-law index between 2500 and 3000~\AA\
($f_{\nu}\propto\nu^{\alpha_{\rm UV}}$; see Brandt et al. 2000).
Vanden~Berk et al. (2001) find that $\langle\alpha_{\rm
UV}\rangle=0.44$.

The flux densities at a rest-frame energy of 2~keV, \fx, without
corrections for intrinsic absorption, were determined from the fits of
the the absorbed power-law model (model~\ref{mod:abspow}) to the
spectra and are listed in Table~\ref{tab:flux}.  These were combined
with \fuv\ from Table~\ref{tab:quasars} to compute the value of \aox,
which we also list in Table~\ref{tab:flux}.  The uncertainty in \aox\
is $\delta\aox = 0.17\; \left[(\delta \fx/\fx)^2 + (\delta
\fuv/\fuv)^2\right]^{1/2}$, where $\delta \fx$ and $\delta \fuv$ are,
respectively, the uncertainties in the X-ray and UV fluxes, resulting
from measurement errors or variability of the source (since the UV and
X-ray observations were not simultaneous).  In the absence of
absorption, the measurement error in \fx\ is dominated by the
uncertainty in the photon index and under this condition it is given
by $\delta\fx / \fx = 0.69 \;\delta\Gamma$, where $\delta\Gamma$ is
the uncertainty in $\Gamma$. For a typical value of
$\delta\Gamma=0.2$, $\delta\fx / \fx = 0.13$, which contributes an
uncertainty of approximately 1--2\% to \aox. Unfortunately, we cannot
assess the contribution of variability to the uncertainty in \aox\
because of the lack of systematic monitoring data, however, we do note
that changes of order unity in either the X-ray or the UV flux change
\aox\ by an amount of order 0.1.

Intrinsic absorption, especially in the UV, is the largest
source of uncertainty in the value of \aox. This uncertainty arises
because we are not able to assess the dust content of the absorber. On
one hand, the presence of low- and intermediate-ionization absorption
lines in the rest-frame UV spectra of 3/4 quasars with intrinsic NALs
suggests that the the ionization state may be low enough for dust to be
present. More specifically, the $\voff=-24,195\;\kms$ system in
Q1700$+$6416 shows the \ion{Si}{2} $\lambda$1260 and \ion{C}{2}
$\lambda$1335 lines (I.P. of 16.3~eV and 24.4 eV, respectively; both
lines are observed in the interstellar medium), while the
$\voff=-21,388\;\kms$ system in Q1107$+$4847 shows the \ion{C}{2}
$\lambda$1335 line (Misawa et al. 2007b). The spectrum of
HS1603$+$3820 presented by Dobrzycki et al. (2007) shows a weak
\ion{Mg}{2} $\lambda$2800 doublet (I.P. of 15.0~eV) in the mini-BAL
system at $\voff\approx -9,000\;\kms$ (the spectra presented in Misawa
et al. 2007a,b do not cover the spectral region of the \ion{Mg}{2}
doublet in any of these quasars). On the other hand, the values of
\aox\ of the quasars with intrinsic NALs in our sample are the same as
those of normal quasars (see the comparisons later in this section).
Since UV extinction by dust can affect \aox\ significantly (as we
detail below), this result suggests that UV extinction in quasars with
intrinsic NALs is not higher than in normal quasars.

We discuss the effects of absorption in detail in the Appendix where
we also emphasize the uncertainties and outline a scheme for applying
corrections. If the dust-to-gas ratio in the intrinsic absorber is
similar to that found in the ISM of the Milky Way, intrinsic UV
absorption could have a dramatic effect on the value of
\aox. Specifically, absorption through an intrinsic column of $1\times
10^{22}$~\cmm, would change \aox\ by 1.5 [see
equation(\ref{eq:naox})]. However, the gas-to-dust ratio in intrinsic
absorbers is extremely uncertain, thus corrections to \aox\ based
solely on the absorbing column measured from X-ray spectra are rather
unreliable. A good case in point is provided by Q0014+8118, the only
quasar in our sample with a measurable intrinsic column density.  The
rest-frame 2~keV flux density, corrected for intrinsic absorption, is
$\fxpr= 179$~nJy. An analogous correction to the rest-frame 2500~\AA\
flux density, assuming the column inferred from the X-ray spectrum and
applying equation~(\ref{eq:dust}), gives $\fuvpr= 25.0$~Jy, leading to
$\aoxpr=-3.12$, which is unreasonable.  We return to this case and
discuss it further in \S\ref{sec:case}.

\begin{figure}
\centerline{\includegraphics[width=3.2in]{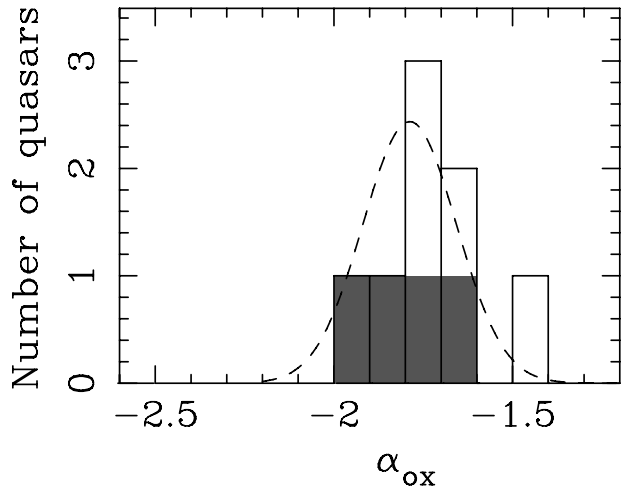}}
\caption{The distribution of values of \aox, evaluated without
  corrections for intrinsic absorption. Shaded bins represent quasars
  with intrinsic NALs, while open bins represent ones without.  For
  the two quasars that were observed more than once (Q1422$+$2309 and
  Q1700$+$6416), we plot a single, representative value of \aox\ (see
  discussion in \S\ref{sec:xuv} of the text).  The dashed line shows
  the distribution of \aox\ for optically bright quasars that are
  similar to those of our sample (Steffen et
  al. 2006). \label{fig:histaox}}
\end{figure}

In Figure~\ref{fig:histaox} we show the distribution of values of
\aox\ among the quasars in our sample, separating the quasars with
intrinsic NALs from those without. For two quasars observed more than
once, Q1422$+$2309 and Q1700$+$6416, we plot only a single,
representative value of \aox; we choose the value obtained from the
deepest X-ray exposure, which happens to be the last observation of
each object. In the same figure, we overplot the distribution of \aox\
among optically bright quasars, as reported by Steffen et al. (2006).
This comparison suggests that there is no discernible
difference between the quasars in our collection hosting intrinsic
NALs and other quasars of comparable luminosity and redshift.

\begin{figure}
\centerline{\includegraphics[width=3.4in]{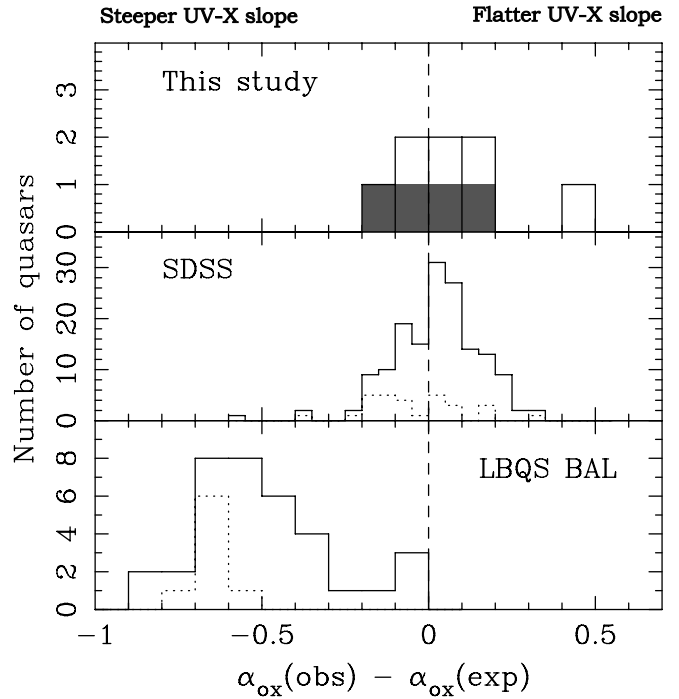}}
\caption{Distribution of \daox, the difference between the observed
  value of \aox\ and the value predicted for that monochromatic UV
  luminosity by the correlation of Steffen et al. (2006). Negative
  values of \daox\ indicate a steeper slope than expected.
  Histograms drawn as dotted lines indicate upper limits. {\it Top
  Panel:} The distribution of \daox\ among quasars in our sample
  (without corrections for intrinsic absorption). Shaded bins
  represent quasars with intrinsic NALs, while open bins represent
  ones without.  For the two quasars that were observed more than once
  (Q1422$+$2309 and Q1700$+$6416), we plot a single, representative
  value of \aox\ (see discussion in \S\ref{sec:xuv} of the text). {\it
  Middle Panel:} The distribution of \daox\ among SDSS quasars from
  Steffen (2006). {\it Bottom Panel:} The distribution of \daox\ among
  BAL quasars from the LBQs (from Gallagher et
  al. 2006). \label{fig:daox}}
\end{figure}

An alternative way of making this comparison is shown in
Figure~\ref{fig:daox}, where we plot the distribution of the
difference between the observed values of \aox\ and that expected
based on the quasar's monochromatic UV luminosity, $\daox=\aox(obs) -
\aox(exp)$ (the values of \daox\ are included in column 6 of
Table~\ref{tab:flux}). More specifically, Strateva et al. (2005) and
Steffen et al. (2006) find a correlation between the monochromatic
luminosity density at 2500~\AA, \luv, and \aox\ for a large sample of
radio-quiet quasars, spanning a wide redshift range. We adopt the form
that includes a weak dependence on redshift, given in equation (5) of
Steffen et al. (2006) as $\aox(exp) = -0.126\;\log\luv -0.01\;z +
2.311$ (the UV-to-X-ray slope becomes steeper as the UV luminosity
increases). A negative value of \daox\ indicates that the UV-to-X-ray
slope is steeper than expected, which would result either from a
suppressed X-ray flux or an enhanced UV flux.  With this in mind, we
plot in Figure~\ref{fig:daox} the distribution of \daox\ among our
quasars (without any intrinsic absorption corrections to \luv) and we
compare it with the corresponding distribution in the Steffen et
al. (2006) sample (top and middle panels, respectively). The two
distributions appear very similar, reinforcing our conclusion that
there is no discernible difference between the X-ray and UV properties
of quasars with and without intrinsic NALs. The only quasar in our
sample that appears to be an outlier in the top panel of
Figure~\ref{fig:daox} is Q0014+8118. This is, in fact, a radio-loud
quasar, whose absolute value of \aox\ is expected to be systematically
smaller than those of radio-quiet quasars (see, for example, Wilkes et
al. 1994).

\begin{figure}
\centerline{\includegraphics[width=3.4in]{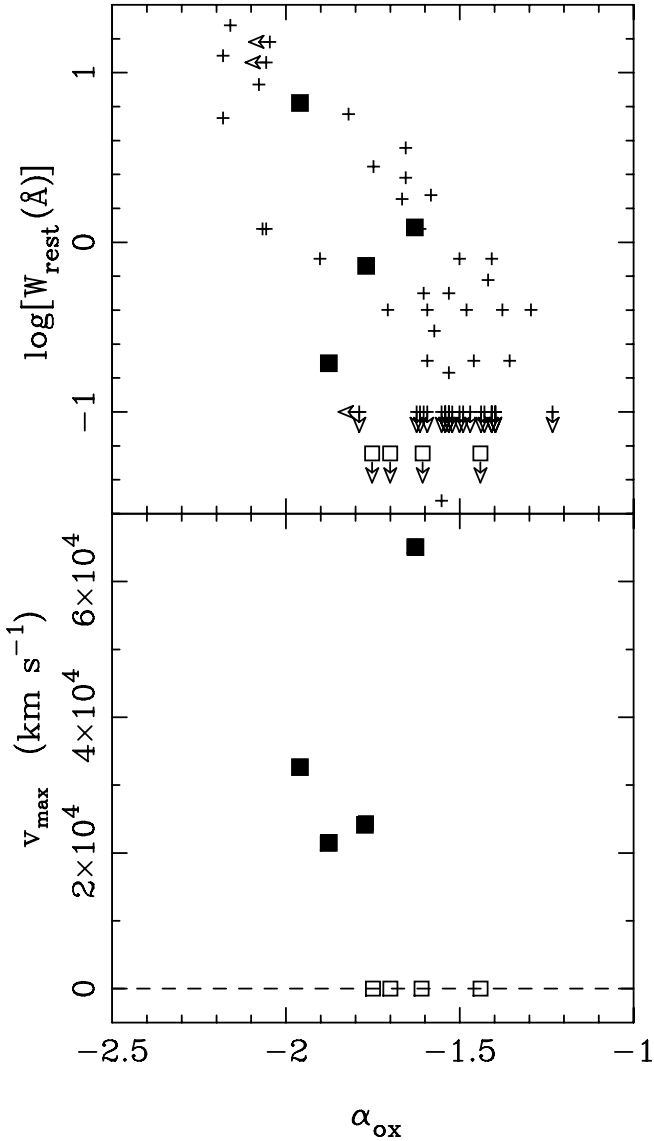}}
\caption{Properties of intrinsic \ion{C}{4} NALs of quasars in our
  sample plotted against \aox\ (evaluated without corrections for
  intrinsic absorption). Filled squares represent quasars with
  intrinsic NALs, while open square represent ones without. For the
  two quasars that were observed more than once (Q1422$+$2309 and
  Q1700$+$6416), we plot a single, representative value of \aox\ (see
  discussion in \S\ref{sec:xuv} of the text). {\it Top Panel:}
  Variation of rest frame equivalent width with \aox.  Their rest
  frame equivalent width is the sum of equivalent widths of all
  intrinsic NALs in the same quasar. The rest-frame equivalent width
  assigned to quasars without intrinsic NALs corresponds to the
  typical observed-frame detection limit of 0.056~\AA\ (see Misawa et
  al 2007b). We adopt a single value of \aox\ for each quasar as
  discussed in in \S\ref{sec:xuv} of the text. The crosses represent
  the associated \ion{C}{4} NALs measured in low-redshift quasars by
  Brandt et al. (2000). {\it Bottom Panel:} Variation of the maximum
  NAL velocity (set to zero for quasars without intrinsic NALs) with
  \aox.
  \label{fig:compaox}}
\end{figure}

We extend our comparison of UV and X-ray properties by plotting the
variation of properties of \ion{C}{4} NALs (rest equivalent width,
\wrest\footnote{Here we use the total equivalent width, summed over
all the intrinsic NALs in the same quasar.}, and maximum blueshift
velocity, \vmax) with \daox. In this context, we take \daox\ to be an
indicator of X-ray absorption, under the assumption that the UV flux
is unabsorbed -- this amounts to neglecting the second term in
equation~(\ref{eq:coraox}). In Figure~\ref{fig:compaox} we plot
\wrest\ and \vmax\ against \daox. In the former plot we overplot the
data points describing nearby PG quasars from Brandt et
al. (2000)\footnote{The \ion{C}{4} NALs in the sample of Brandt et
al. (2000) are associated to the quasars but they are not necessarily
intrinsic.  We converted the values of \aox\ in Brandt et al. (2000),
evaluated using the 3000~\AA\ UV flux, to the convention used here, as
described at the beginning of \S\ref{sec:xuv}.} to show that the
quasars in our sample with intrinsic NALs follow the general
anti-correlation between \wrest\ and \aox.  However, since our sample
is rather small, this result requires verification using a
considerably larger sample of quasars with intrinsic NALs.  In the
latter plot, there is no obvious trend between \vmax\ and \aox; we
return to this issue in \S\ref{sec:discussion}, where we compare
intrinsic NALs to BALs and place them in the broader context of
intrinsic absorption lines in quasars.

\goodbreak
\section{Summary of Results and Discussion}\label{sec:discussion}

The results presented above show that there are no differences in
X-ray properties between quasars with and without intrinsic NALs in
our sample. The spectra of all the quasars in our sample can be
described by a very simple model, namely a power law modified by
absorption by nearly-neutral matter at the redshift of the source.  In
all but one quasar, we are only able to obtain upper limits to the
column density of the intrinsic X-ray absorber; these limits
are of order a few $\times 10^{22}$~\cmm.  The possibility that the
intrinsic X-ray absorbers are ionized can be neither confirmed
nor ruled out based on the data we have analyzed here. The only quasar
in which we have found a measurable column density is Q0014+8118
($\nh\approx 1\times 10^{22}$~\cmm), which does {\it not} host any
intrinsic NALs. We discuss this quasar further below. Our comparison
with larger samples of quasars of comparable luminosity and redshift
shows that quasars with intrinsic NALs do not differ from the general
population in either their X-ray spectra or their spectral energy
distributions as quantified by \aox. The rest-frame equivalent widths
of intrinsic NALs follow the same trend with \aox\ as nearby quasars
and Seyfert galaxies, suggesting a relation between the medium
responsible for the intrinsic NALs and the medium responsible for the
X-ray continuum absorption. However, X-ray observations of a larger
sample of quasars with intrinsic NALs is sorely needed to strengthen
our conclusions.  A conclusion that follows immediately from the above
results is that outflows that manifest themselves in the form of
intrinsic NALs could be present in all quasars. We showed in our
earlier work that intrinsic NALs are very common (occurring in 50\% of
all quasars at $z\sim 2$--4; Misawa et al. 2007b) and the results
presented here indicate that the typical spectral characteristics of
quasars in general do not preclude their hosting intrinsic NALs.

\begin{figure}
\centerline{\includegraphics[width=3.4in]{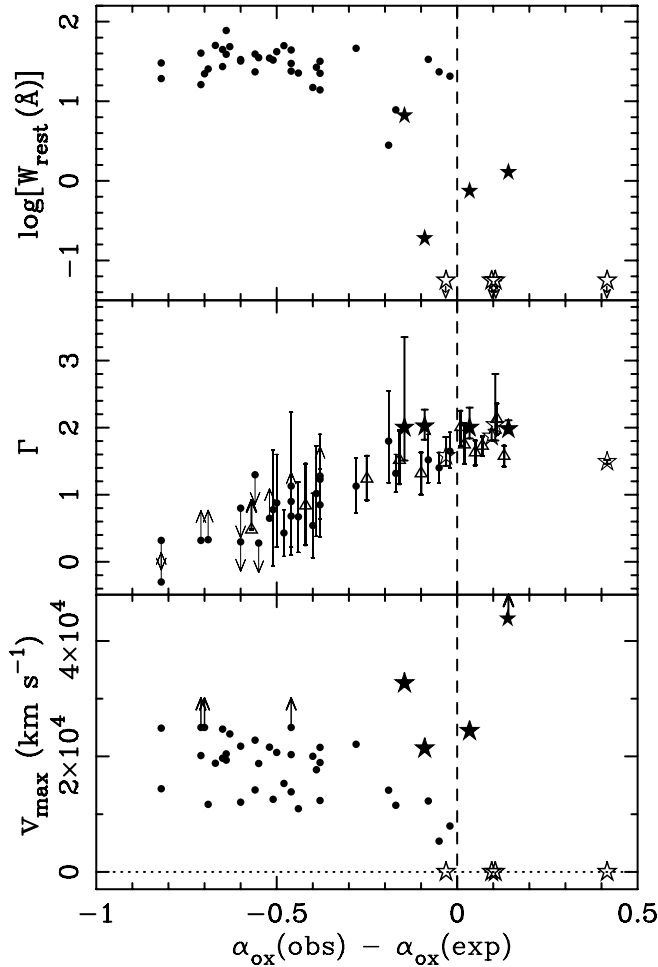}}
\caption{Comparison of the properties of quasars in our sample with
  the properties of BAL quasars. Quasars from our sample with
  intrinsic NALs are represented by stars, while those without are
  represented by open stars. For the two quasars that were observed
  more than once (Q1422$+$2309 and Q1700$+$6416), we plot a single,
  representative value of \aox\ (see discussion in \S\ref{sec:xuv} of
  the text). Filled circles represent BAL quasars from the sample of
  Gallagher et al. (2006).  {\it Top Panel:} Variation of the total
  rest-frame equivalent width (summed over all intrinsic NALs in the
  same quasar) with \daox. The rest-frame equivalent width assigned to
  quasars without intrinsic NALs corresponds to the typical
  observed-frame detection limit of 0.056~\AA\ (see Misawa et al
  2007b). {\it Middle Panel:} Variation of X-ray photon index with
  \daox.  Open triangles in this panel represent the extremely red
  quasars from Hall et al. (2006). {\it Bottom Panel:} Variation of
  maximum blueshift velocity of intrinsic NALs with \daox. One
  intrinsic NAL quasar from our sample (Q0130$-$4021; $\vmax\approx
  65,000~\kms$) is off scale and marked with upward
  arrow.\label{fig:correlations}}
\end{figure}

In the remainder of this section, we compare the properties of quasars
with intrinsic NALs from our sample with those of BAL quasars studied
by other authors, we use our results to constrain scenarios for the
geometry and location of the intrinsic NAL gas, and we discuss the
properties of the absorber in Q0014+8118. We conclude by noting open
questions and possible directions for future work.

\subsection{Comparison with BAL quasars}\label{sec:comp}

We compare the properties of the NAL quasars in our sample with those
of BAL quasars in the sample of Gallagher et al. (2006) and very red
quasars from Hall et al. (2006); their redshifts are between 1.4 and
2.8 and their luminosities are comparable to those of our quasars.  In
Figure~\ref{fig:daox}, we show the distribution of {\it measured}
values of \daox\ among BAL quasars on the same scale as the
distribution of \daox\ among quasars in our sample. There is an
obvious difference between the two distributions; a Kolmogorov-Smirnov
test yields a chance probability of 0.003 that the intrinsic NAL
quasars in our sample and the BAL and very red quasars from the bottom
panel of Figure~\ref{fig:daox} were drawn by chance from the same
parent population.  This suggests that the X-ray source in intrinsic
NAL quasars is not as heavily obscured as in BAL quasars (see the
discussion of the \daox\ distribution in BAL quasars by Gallagher et
al. 2006).

In Figure~\ref{fig:correlations}, we plot properties of UV BALs as
well as the X-ray photon index for the Gallagher et al. (2006) sample
against \daox. In the same plots we include the quasars from our own
sample for comparison. The top two panels of
Figure~\ref{fig:correlations} suggest that quasars with intrinsic NALs
fit in a progression between BALs and ``normal'' quasars. As the
rest-frame equivalent width of the \ion{C}{4} absorption line
decreases, \daox\ increases, in a manner that is qualitatively
consistent with a decrease in the column density of the absorber. This
picture is supported by the photon index vs \daox\ plot, where quasars
with intrinsic NALs occupy the low-column-density end of the BAL
distribution. It is reassuring that the mini-BAL quasar in our sample,
HS1603+3820, falls closer to the BAL locus in both of the top two
panels of Figure~\ref{fig:correlations} (as well as in the top panel
of Fig.\ref{fig:compaox}) than the intrinsic NAL quasars.

In contrast, in the plot of maximum blueshifted velocity vs \daox\
shown in the bottom panel of Figure~\ref{fig:correlations}, quasars
with intrinsic NALs do not fit into the BAL sequence, even though this
sequence appears to connect smoothly to the ``normal'' quasars. This
behavior can be reconciled with the behavior seen in the top two
panels of the same figure in the context of an equatorial wind
scenario. In the top two panels of Figure~\ref{fig:correlations} the
BAL quasar tracks are defined primarily by the column density of the
absorbing medium since $\log\wrest$, $\Gamma$, and \daox\ are all measures
of the column density (the values of $\Gamma$ of BAL quasars were
inferred from the hardness ratio, thus they are directly affected by
absorption and so are the 2~keV flux densities that are based on the
same data). On the other hand, the orientation of the outflow relative
to the line of sight and the acceleration mechanism play an important
role in defining the BAL quasar track in the bottom panel of this
figure, since they influence the value of \vmax\ (see for example, the
discussion in Gallagher et al 2006). Therefore, the deviation of
intrinsic NAL quasars from the BAL track suggests that the NAL gas is
not part of the dense BAL flow thought to be located near the base of
the wind, although it may still be associated with it. In other words,
our line of sight through the NAL gas does not generally pass through
the dense BAL gas. This conclusion also serves as a test of scenarios
for the location of the NAL gas within the greater outflow, as we
explain further in \S\ref{sec:geom}, below.

\subsection{Geometry and Location of Absorbing Medium}\label{sec:geom}

Using our observational results, we can test and constrain scenarios
for the location of the intrinsic NAL gas in an (equatorial)
accretion-disk wind whose dense, low-latitude parts are likely to give
rise to BALs. The two competing scenarios are depicted in Figures~1--3
of Elvis (2000) and in Figure~13 of Ganguly et al. (2001). Elvis
(2000) proposed that quasars with intrinsic NALs are viewed at a large
inclination angle relative to the disk/wind axis (larger than in the
case of BALs).  As a result the column density through the NAL line of
sight is $\nh\gs 10^{22}$~\cmm\ and the apparent blueshift of the
intrinsic NALs should be considerably smaller than that of BALs,
($\voff\sim10^3$~\kms). On the other hand, in the Ganguly et
al. (2001) picture the NAL gas is located {\it above} the BAL gas,
closer to the axis of the disk/wind. The column density is lower
because the NAL gas is distributed in small parcels and the blueshifts
can be as high as those found in BALs (or even higher).

Our results favor the Ganguly et al. (2001) scenario for a number of
reasons. First, we do not detect large absorbing column densities; we
find that $\nh\ls {\rm a\; few}\times 10^{22}$~\cmm\ in all of our
quasars. Second, the observed blueshifts of intrinsic NALs are
comparable to or higher than those of BALs, suggesting that the
absorbing gas is not associated with the base of the wind. Rather,
this gas should have traveled an appreciable distance from its launch
point and attained a speed close to its terminal speed (see, for
example, Murray et al. 1995 and Hamann 1998). Third, the comparison of
intrinsic NALs and BALs presented above (\S\ref{sec:comp} and bottom
panel of Fig.~\ref{fig:correlations}) shows that the intrinsic NALs do
not connect smoothly to the BAL sequence in the \vmax--\daox\ diagram,
suggesting that the NAL gas is not a part of the dense BAL flow.  In
contrast, the Ganguly et al. scenario passes these tests and also
places the NAL gas far enough away from the central continuum source
so that (a) it is not highly ionized, and (b) it can be accelerated to a
high speed.

\subsection{The Case of Q0014+8118}\label{sec:case}

Q0014+8118 is the only quasar in which we have found excess absorption
but it is also a quasar not known to have intrinsic UV NALs. We
explore here whether these two observational results can be
reconciled.  We have re-examined the rest-frame UV spectrum of this
quasar presented in Misawa et al. (2007b) and found no strong
absorption lines near the redshift of this quasar, which could be
attributed to a neutral absorber with the column density comparable to
that measured in the X-ray spectrum. We did, however, find an
intervening absorption line system at $z=1.1$ with strong
\ion{Fe}{2}$\;\lambda$2600, \ion{Mg}{2}$\;\lambda\lambda$2796,2803 and
\ion{Mg}{1}$\;\lambda$2853 absorption lines with the following
rest-frame equivalent widths: $W_{rest}(\lambda 2600)=2.06\;$\AA,
$W_{rest}(\lambda 2796)=3.23\;$\AA, and $W_{rest}(\lambda
2853)=0.67\;$\AA. This raises the possibility that the excess X-ray
absorption is associated with the lower-redshift intervening system.
We consider each of these two scenarios in turn.

To assess whether the X-ray absorber could be intrinsic to the quasar,
we have carried out a number of photoionization simulations using the
code Cloudy (Ferland et al. 1998) adopting the Mathews \& Ferland
(1987) model for the quasar spectral energy distribution (but modified
to match the X-ray photon index of Q0014+8118), a Solar metallicity
and abundance pattern, the total Hydrogen column density determined
from the X-ray spectrum, and two different values of the density
($10^4$ and $10^8$~\cmmm; the results turn out not be sensitive to
the density). We computed four models with ionization parameters
corresponding to $\log\xi_{\rm XSTAR}=1.04$, 1.84, 2.24, and 2.54,
which span the range of values inferred for the X-ray absorber (see
\S\ref{sec:notes}). In all cases, the equivalent widths of the
\ion{C}{4}, \ion{N}{5}, and \ion{O}{6} absorption lines are large
enough that these lines should have been detected in the
high-resolution spectrum presented in Misawa et al. (2007b). Missing
these lines would require a rather unlikely set of circumstances,
namely: the ionization parameter should be at the high end of the
allowed range so that the lines are weak and the outflow velocity
should be such that the \ion{C}{4} line is out of the range of the
observed UV spectrum, the \ion{N}{5} line should fall in the gap
between echelle orders, and the \ion{O}{6} line should be hidden in
the Ly$\alpha$ forest. Therefore, we disfavor the intrinsic
interpretation of the absorber.

To assess the intervening absorber hypothesis we make use of the
statistical observational results of Rao, Turnshek, \& Nestor (2006).
According to these authors, an intervening absorber with the observed
\ion{Fe}{2}, \ion{Mg}{2}, and \ion{Mg}{1} equivalent widths has a 65\%
probability of being a damped Ly$\alpha$ system (DLA), with an average
hydrogen column density of $4\times 10^{20}$~\cmm\ and a column
density of $1\times 10^{21}$~\cmm\ being reasonably likely. Moreover,
the observed relative strengths of the \ion{Fe}{2} and \ion{Mg}{2}
lines suggest that the hydrogen column density can plausibly be as
high as $5\times 10^{21}$~\cmm, while the observed \ion{Mg}{1}
strength suggests a 50\% probability of $N_{\rm H} > 1\times
10^{21}$~\cmm.  Thus we favor the intervening absorber hypothesis
because the known intervening system at $z=1.1$ has a high probability
of producing just the X-ray absorption we observe. In this context we
may also understand the low attenuation of the rest-frame UV light of
this quasar. By comparing the colors of large samples quasars with and
without DLAs, Ellison, Hall, \& Lira (2005) find that the dust
content of DLAs is very small and place a limit of $E(B-V)<0.04$ on
the color excess that they produce.

\subsection{Open Questions and Future Prospects}\label{sec:prospects}

There are a number of outstanding questions that our exploratory
survey was not able to address.

\begin{itemize}

\item
We have not detected an ionized absorber in any of the quasars with
intrinsic NALs, even though such features are common in the X-ray
spectra of Seyfert galaxies. Contributing factors to this were the low
$S/N$ of the X-ray spectra and the high redshift of the quasars
themselves (the signature of a warm absorber is most pronounced at
energies that are redshifted out of the observable band). Detecting or
placing limits on a warm absorbing medium is important because it
provides a test of models for the structure of the outflow (especially
the ionization structure at high latitudes above the disk; see, for
example, Proga et al. 2000). Moreover, a warm medium has been invoked
to interpret the variability of mini-BALs (see the discussion of
HS1603+3820 by Misawa et al. 2007a).

\item
We have not detected high-velocity X-ray absorption lines in any of
our spectra. This is not a surprise since such lines are rare. More
specifically, unresolved lines have been found in about half a dozen
Seyfert galaxies and quasars at $z<0.1$, including MCG-5-23-16 (Braito
et al. 2006), IC~4329a (Markowitz et al. 2006), IRAS~13197-1627
(Dadina \& Cappi 2004), Mrk~509 (Dadina et al. 2005), and PG~1211+143
(Pounds et al. 2007).  The observed outflow velocities are of order
$0.1\; c$, while the rest-frame equivalent widths typically range
between 10 and 100~eV.  Broader, and somewhat stronger absorption
lines have also been detected in a handful of high-redshift quasars
($z\sim 2$--4), namely PG~1115+080 (Chartas et al. 2003, 2007a),
H~1413+117 (Chartas et al 2007b), and APM~08279+5255 (Chartas et
al. 2002; Hasinger et al. 2002).  In these cases the outflow speeds
are of order a few tenths of the speed of light and the rest-frame
equivalent widths range between a few hundred eV and a few keV. In
addition to the cosmological implications of high-velocity X-ray
absorption lines noted in \S\ref{sec:intro}, their detection will also
help us constrain the ionization state of the NAL gas. The limits we
have set for Q0014+8118 are rather stringent ($< 28$~eV), comparable
to the weakest lines ever detected. In other quasars in our sample the
limits are considerably higher, thus uninteresting, owing to the worse
$S/N$ in their X-ray spectra. Nevertheless, it is interesting that the
absence of high-velocity X-ray absorption lines in Q0014+8118
coincides with the absence of intrinsic UV NALs.

\item
Our results, especially Figure~\ref{fig:correlations}, raise the issue
of whether the UV and X-ray properties of NAL quasars connect smoothly
to those of BAL quasars. Related to this issue is the question of
where do mini-BALs fall in the grand scheme of things. Even though we
have found some tantalizing trends, our sample of objects is too small
for a firm conclusion on these issues. In this respect, further
observations of mini-BALs would be particularly useful because they
may fill in the gap between NALs and BALs in
Figure~\ref{fig:correlations}.

\end{itemize}

The above questions can be addressed by employing two complementary
strategies in future observations. Long exposures of selected objects
can yield high-$S/N$ X-ray spectra that can be used for a sensitive
search for warm absorbers or high-velocity X-ray NALs. At the same
time, (relatively shallow) observations of larger samples of quasars
hosting NALs and especially mini-BALs will allow us to explore
connections between the corresponding quasar populations.

\acknowledgments 

We thanks the anonymous referee for helpful comments and suggestions.
We acknowledge support from NASA grant NAG5-10817.  This work
was also partially supported by the Sumitomo foundation (070380).  We
have made use of the NASA/IPAC Extragalactic Database (NED) which is
operated by the Jet Propulsion Laboratory, California Institute of
Technology, under contract with the National Aeronautics and Space
Administration.

\clearpage 


\appendix

\section{Relation Between UV and X-Ray Extinction and Effect 
on the Value of \aox.}

The observed UV flux is attenuated in the rest frame of the quasar
according to $\fuvobs = \fuvintr\; 10^{-0.4\;\sAuv}$, where $\Auv$ is
the extinction at 2500~\AA\ in magnitudes. After passing through the
ISM of the Milky Way, the quasar UV light is attenuated further by a
factor of $10^{-0.4\; A_{\rm \lambda(1+z)}}$ (where
$\lambda=2500$~\AA\ and $z$ is the redshift of the quasar). However,
for quasars at $z\gs 2$, the latter factor is negligible compared to
the former because $\lambda(1+z)$ falls in the near-IR band and
$A_{\rm IR} \ll A_{\rm UV}$. The extinction law of Seaton (1979) gives
\begin{equation}
{\Auv\over E(B-V)} - {A_{\rm V}\over E(B-V)} = 2.5 \quad\Rightarrow\quad
\Auv = A_{\rm V} + 2.5\; E(B-V) = 5.7\; E(B-V)\; ,
\end{equation}
where $A_{\rm V} = 3.2\; E(B-V)$ is the visual extinction and
$E(B-V)\equiv A_{\rm B}-A_{\rm V}$ is the color excess or
``reddening'' ($A_{\rm B}$ is the extinction in the B-band).  All
extinction laws currently available (including the SMC law, which is
thought to describe the extinction in the UV spectra of quasars at
$z<4$; see Hopkins et al. 2004) are very similar at $\lambda \ge
2500$~\AA. This is illustrated in Figure~1 of Pr\'evot et al. (1984)
and in Figure~1 of Calzetti, Kinney, \& Storchi-Bergmann (1994).

Similarly, the X-ray flux is attenuated according to: $f^{\;\rm
obs}_{\rm 2~keV} = f^{\;\rm intr}_{\rm 2~keV}\; e^{-N_{\rm H}\;
\sigma_{\rm 2\;keV}}$, where $N_{\rm H}$ is the equivalent hydrogen
column density and $\sigma_{\rm 2\;keV}$ is the photoelectric
absorption cross-section at 2~keV. According to Morrison \& McCammon
(1983), $\sigma_{\rm 2\;keV}=3.19\times 10^{-23} ~{\rm cm}^{-2}$,
determined by the abundance of Mg and Si (both $\alpha$-elements). We
note that Morrison \& McCammon (1983) adopt the elemental abundances
of Anders \& Ebihara (1982), while different authors (namely, Anders
\& Grevesse 1989; Feldman 1992; Grevesse \& Sauval 1998) derive
abundances for these two elements that are within $\approx 5$\% of
that value. A potential cause for concern are the high metal
abundances observed in the immediate vicinity of quasar central
engines (see Hamann \& Ferland 1999 for a review). More specifically,
Hamann et al. (2002) find that $\alpha$-elements, such as C and O, are
overabundant in the broad-emission line regions of quasars by a factor
of $\sim 3$ relative to the Sun. This will likely affect the X-ray
attenuation at 2~keV, but the more important question is whether it
will also affect the UV attenuation in the same manner. The answer to
this question is unclear as it depends on whether the higher metal
abundance in that gaseous phase is also accompanied by a similarly
higher abundance of dust grains, which are responsible for the UV
extinction.

Inserting the expressions for the {\it observed} fluxes into the
definition of \aox\ in equation~(\ref{eq:defaox}), we obtain
\begin{equation}
\alpha^{\rm obs}_{\rm ox} - \alpha^{\rm intr}_{\rm ox} = 0.384
\left(- N_{\rm H}\; \sigma_{\rm 2\;keV}\; \log e + 0.4\; \Auv\right) 
= -0.053\; N_{22} + 0.88\; E(B-V)\; ,
\label{eq:coraox}
\end{equation}
where $N_{22}=N_{\rm H}/10^{22}\;{\rm cm}^{-2}$. The first term on the
right-hand side of equation~(\ref{eq:coraox}) represents the effect of
X-ray attenuation, while the second term represents the effect of UV
extinction.  The relative importance of these two terms can be
assessed by noting that in the Milky Way the dust-to-gas ratio is such
that
\begin{equation}
N_{\rm H}/A_{\rm V} = 1.79\times 10^{21}~{\rm cm^{-2}~mag^{-1}}
\quad\Rightarrow\quad
E(B-V) = 1.7\; N_{\rm 22}~{\rm mag}
\label{eq:dust}
\end{equation}
(Predehl \& Schmitt 1995). Therefore the second term on the right-hand
side of equation~(\ref{eq:coraox}) (UV extinction) is aproximately 30
times more important than the first term (X-ray attenuation), if the
gas-to-dust ratio is comparable to that in the Milky Way. If we insert
equation~(\ref{eq:dust}) into equation~(\ref{eq:coraox}), we get
\begin{equation}
\alpha^{\rm obs}_{\rm ox} - \alpha^{\rm intr}_{\rm ox} = 1.5\; N_{22}\; . 
\label{eq:naox}
\end{equation}
Equation~(\ref{eq:naox}) relies on the assumption that the value of
$N_{\rm H}/A_{\rm V}$ (determined by the dust-to-gas mix) in the
vicinity of the quasar is the same as that in the Milky Way. This need
not be true in general. It is possible, for example, that the
intrinsic absorber is close enough to the quasar continuum source that
the intense radiation field destroys the dust grains (e.g., Netzer \&
Laor 1993). Another possibility is that the dust grain distribution in
the absorber is different from that in the Milky Way, as is the case
with very-high redshift quasars (at $z>4$; see Maiolino et al. 2006
and references therein). Therefore, equation~(\ref{eq:coraox}) is
preferable to equation~(\ref{eq:naox}), as long as $E(B-V)$ can be
determined independently from $N_{\rm H}$. An illustrative example of
the dangers of applying equation~(\ref{eq:naox}) without due caution
is provided by the quasar Q0014+8118, discussed in \S\ref{sec:xuv}.
Even though a significant column is detected in absorption in the
X-ray spectrum of this quasar, its rest-frame UV light does not appear
to be significantly attenuated.

\clearpage


\end{document}